\def\MET{\mbox{${\hbox{$E$\kern-0.6em\lower-.1ex\hbox{/}}}_T$}}
\def\RPV{\mbox{${\hbox{$R$\kern-0.6em\lower-.1ex\hbox{/}}}_p$}}
\def\etal{{\sl et al.}}

\documentstyle[12pt,lathuile,graphics,epsfig,amsmath,amssymb]{article}
\begin{document}

\title{Recent Results in Search for New Physics at the Tevatron (Run 1)}
\author{
John Zhou \\
{\em Rutgers University} \\
{\em P.O.Box 500, MS 318,}\\
{\em Batavia, IL 60510, U.S.A.} \\ \\
(on behalf of CDF and D\O\ Collaborations)
}
\maketitle
\baselineskip=14.5pt
\begin{abstract}
We present some new results on searches for new physics at the Tevatron
Run 1 (1992 -- 1996). The topics covered are searches for R-Parity
violating and conserving m{\sc SUGRA}, large extra dimensions in di-photon and
monojet channels, leptoquark in jets~+~\MET\ channel, and two model
independent searches. All results were finalized during the past year.
\end{abstract}

\baselineskip=17pt

\newpage
\section{Introduction}
Tevatron Run I has been a great success for High Energy Physics. For the
period between October 1992 and February 1996, about $120 \; {\rm
pb^{-1}}$ of data were collected by the two competing experiments and
collaborations: CDF and D\O. Both collaborations have made many
important measurements and discoveries with their powerful and
multi-purpose detectors\cite{cdfnim,d0nim}, culminated by the discovery
of the top quark in 1995. They are also engaged in search for new
physics beyond the Standard Model (SM). Though no convincing evidence of
new physics was found, the searches have extended our understanding of
the fundamentals of the universe and have led our quest for ultimate
understanding in a concerted direction.

This paper reports nine results on searches for new physics conducted
recently at the Tevatron by CDF and D\O. We cover the topics of {\sc
SUSY}, large extra dimension, leptoquark, and model independent
searches.

\section{Search for m{\sc SUGRA}}
Minimal supergravity or m{\sc SUGRA}\cite{msugra} is a model which provides a
framework for the spontaneous breaking of the
supersymmetry\cite{susy}. In this model, {\sc SUSY} is broken in the hidden
sector of the theory and this breaking is communicated to the physical
sector of the theory through gravitational interactions. There are five
parameters to completely determine the {\sc SUSY} sector of the theory:

\begin{itemize}
\item $m_{0}$: common scalar particle mass at the {\sc SUSY} breaking scale
$M_{X}$\footnote{$M_{X}$ is usually the GUT Scale ($10^{16} \;
{\rm GeV}$) or the Planck scale ($10^{19} \; {\rm GeV}$).};
\item $m_{1/2}$: common gaugino mass at the $M_{X}$ scale;
\item $A_{0}$: common trilinear coupling at the $M_{X}$ scale;
\item ${\rm tan}\beta$: ratio of the vacuum expectation values of the two
Higgs doublets;
\item ${\rm sign}(\mu)$: $\mu$ is the Higgsino mass parameter.
\end{itemize}

An additional parameter called R-Parity is introduced and is defined as:
$R_{p}=-1^{3B+L+2S}$\cite{r-parity}, where $B$ and $L$ are baryon and
lepton numbers, respectively, and $S$ refers to spin. A superpotential
for {\sc MSSM}, the minimal supersymmetric extension of the standard
model\cite{mssm}, can be written as the following:

\begin{eqnarray}
W & = &
\overline{u}{\rm \bf y_{u}}QH_{u} - \overline{d}{\rm \bf y_{d}}QH_{d} -
\overline{e}{\rm \bf y_{e}}LH_{d} + \mu H_{u}H_{d} + \nonumber \\
& & \lambda_{ijk}L^{i}L^{j}\overline{e}^{k} +
\lambda'_{ijk}L^{i}Q^{j}\overline{d}^{k} +
\lambda''_{ijk}\overline{u}^{i}\overline{d}^{j}\overline{d}^{k} +
\mu'_{i}L^{i}H_{u}.
\label{eqn:superpotential}
\end{eqnarray}

\noindent where the first line describes
$R_{p}$-conserving couplings and the second line describes
$R_{p}$-violating couplings; ${\rm \bf y_{u}}$, ${\rm \bf y_{d}}$, and
${\rm \bf y_{e}}$ are $3 \times 3$ Yukawa coupling matrices; $Q$ and $L$
are left-handed quark and lepton supermultiplets, respectively;
$\overline{u}$, $\overline{d}$, $\overline{e}$ are the right-handed
singlets of the up and down type (s)quarks and (s)leptons, respectively;
$H_{u}$ and $H_{d}$ are the two Higgs doublets; $\mu$ is the Higgsino
mass parameter; $\lambda$, $\lambda'$, and $\mu'$ are the coupling
strengths for lepton number violating interactions and $\lambda''$ is the
coupling strength for baryon number violating interactions.

We describe in this paper four searches for m{\sc SUGRA} under various
additional constraints.

\subsection{CDF RPV m{\sc SUGRA} search in decays of stop pair}
In this analysis, we assume that stop pair are produced through
$R_{p}$-conserving processes and then decay through $R_{p}$-violating
process: $\tilde{t}\overline{\tilde{t}} \rightarrow \overline{\tau}_{l}
+ b + \tau_{h} + \overline{b} + X$, where $\tau_{l}$ and $\tau_{h}$ are
leptonically and hadronically decayed $\tau$, respectively. We also
assume that $\lambda'_{333}$ in Eq.~(\ref{eqn:superpotential}) dominates
the couplings.

The key to this analysis is the identification of $\tau_{h}$. The
following criteria are used to select $\tau_{h}$:

\begin{itemize}
\item $\tau_{h}$ candidates are clusters with $P_{T} > 15 \; {\rm GeV}$
and $|\eta_{det}|<1.0$;
\item Number of tracks and $\pi^{0}$'s in a narrow cone around a $\tau$
cluster candidate are consistent with those coming from a $\tau$;
\item $E/p$ and isolation energy of tracks and reconstructed $tau$ mass are
consistent with those of a $\tau$.
\end{itemize}

\begin{figure}[h!tb] \centering
	\epsfig{file=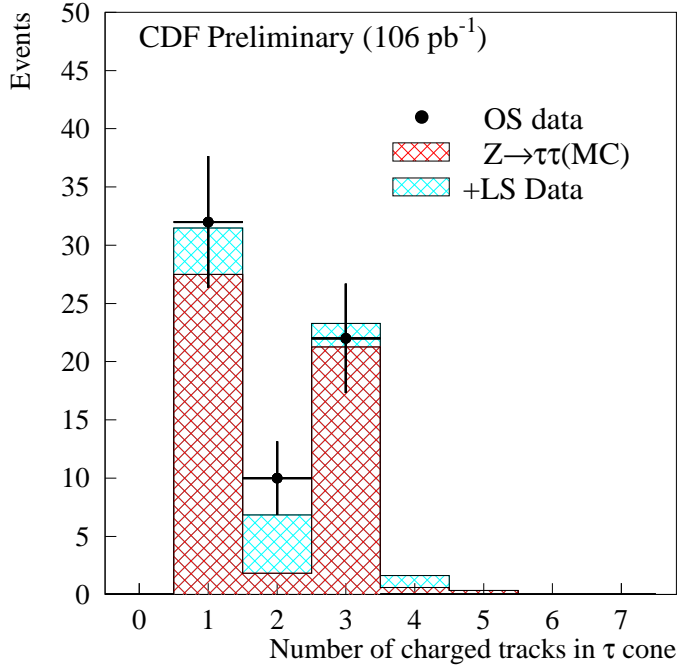, height=10cm}
	\caption{\it Number of tracks in a $\tau$ cone in $Z \rightarrow
	\tau\tau + 0 \; {\rm jet}$ events. Points are data. The darkly
	hatched histogram is the prediction from Monte Carlo, and the
	lightly hatched histogram is the estimation of background from
	data with like-sign charge. Charge of a $\tau$ candidate is the
	sum of track charges in the $\tau$ cone.}
	\label{fig:rpv_stop_echan_tau_ntracks}
\end{figure}

Figure~\ref{fig:rpv_stop_echan_tau_ntracks} shows the number of tracks
in a $\tau$ cone. The 1-prong and 3-prong structures of the $\tau$
candidates are prominent.

A total of $106 \; {\rm pb^{-1}}$ of data are used in this analysis. The
major SM backgrounds come from $Z, \gamma^{*} + {\rm jets}$, Diboson,
$W(e\nu, \mu\nu) + {\rm jets}$, $W(\tau\nu) + {\rm jets}$, and multijet
events. The first two are physics backgrounds which have the same final
states as the signal while the rest are $\tau$ fakes of by jets.

Leptonically decayed $\tau_{l}$ is identified with a tagging electron or
muon. We require $E^{e}_{T}>10 \; {\rm GeV}$, $|\eta^{e}_{det}| < 1.0$
for the electron channel or $P^{\mu}_{T} > 10 {\rm GeV/c}$,
$|\eta^{\mu}_{det}| < 1.0$ for the muon channel. In order to increase
the signal significance, the following additional selection cuts are
applied:

\begin{itemize}
\item transverse mass of the lepton and the \MET: $M_{T}({\rm lepton,
\MET}) < 35 \; {\rm GeV/c^{2}}$;
\item the scalar sum of $E_{T}$ of the lepton, $\tau_{h}$, and \MET: $H_{T}({\rm
lepton}, \tau_{h}, \mbox{\MET}) > 70 \; {\rm GeV}$;
\item ${\rm N_{jet}} \ge 2$ with $E^{\rm {jet}}_{T} > 15 \; {\rm GeV}$.
\end{itemize}

\begin{figure} [h!tb] \centering
\begin{minipage}{0.5\linewidth}
  \centering\epsfig{file=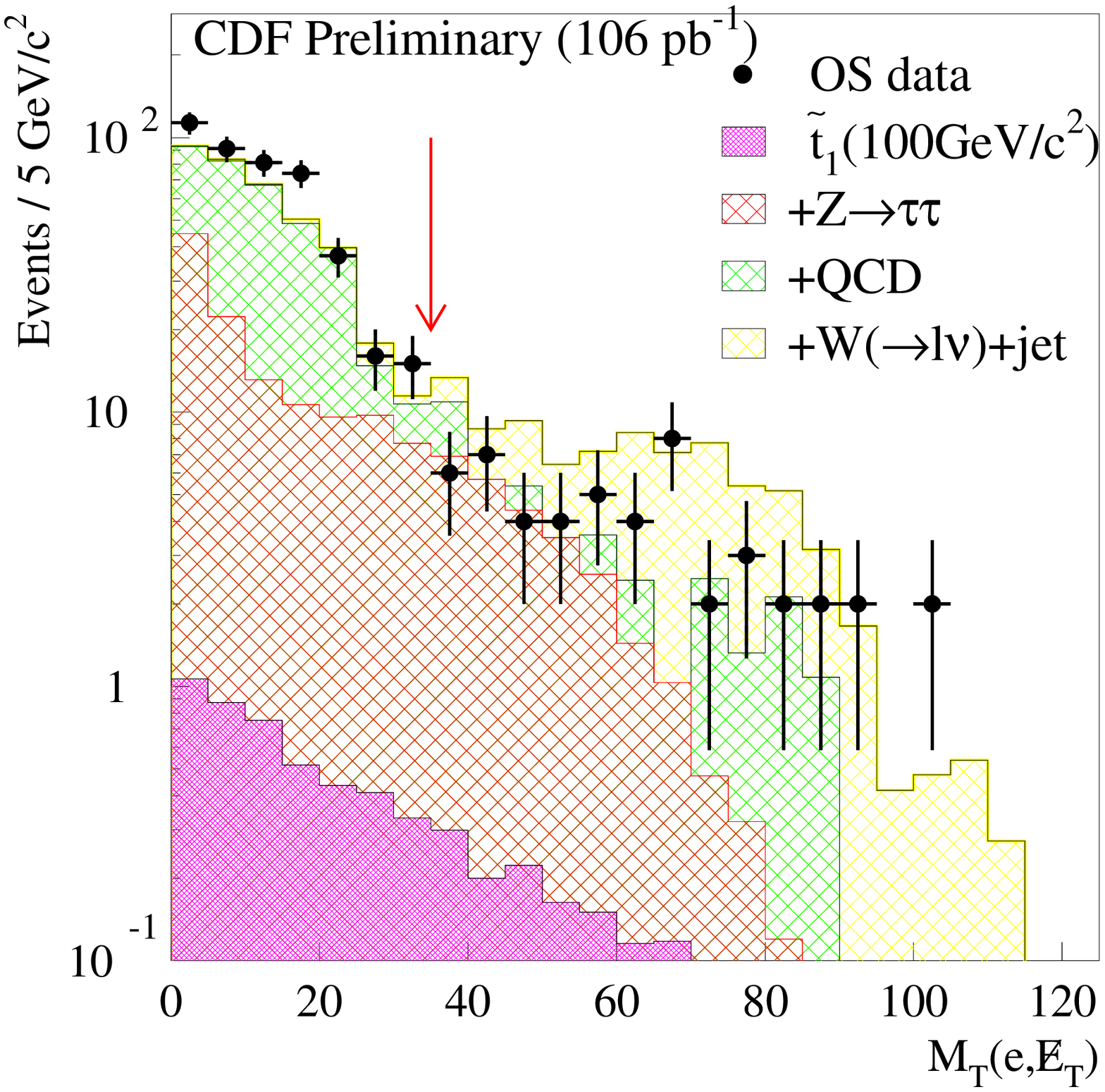,width=\linewidth}
\end{minipage}\hfill
\begin{minipage}{0.5\linewidth}
  \centering\epsfig{file=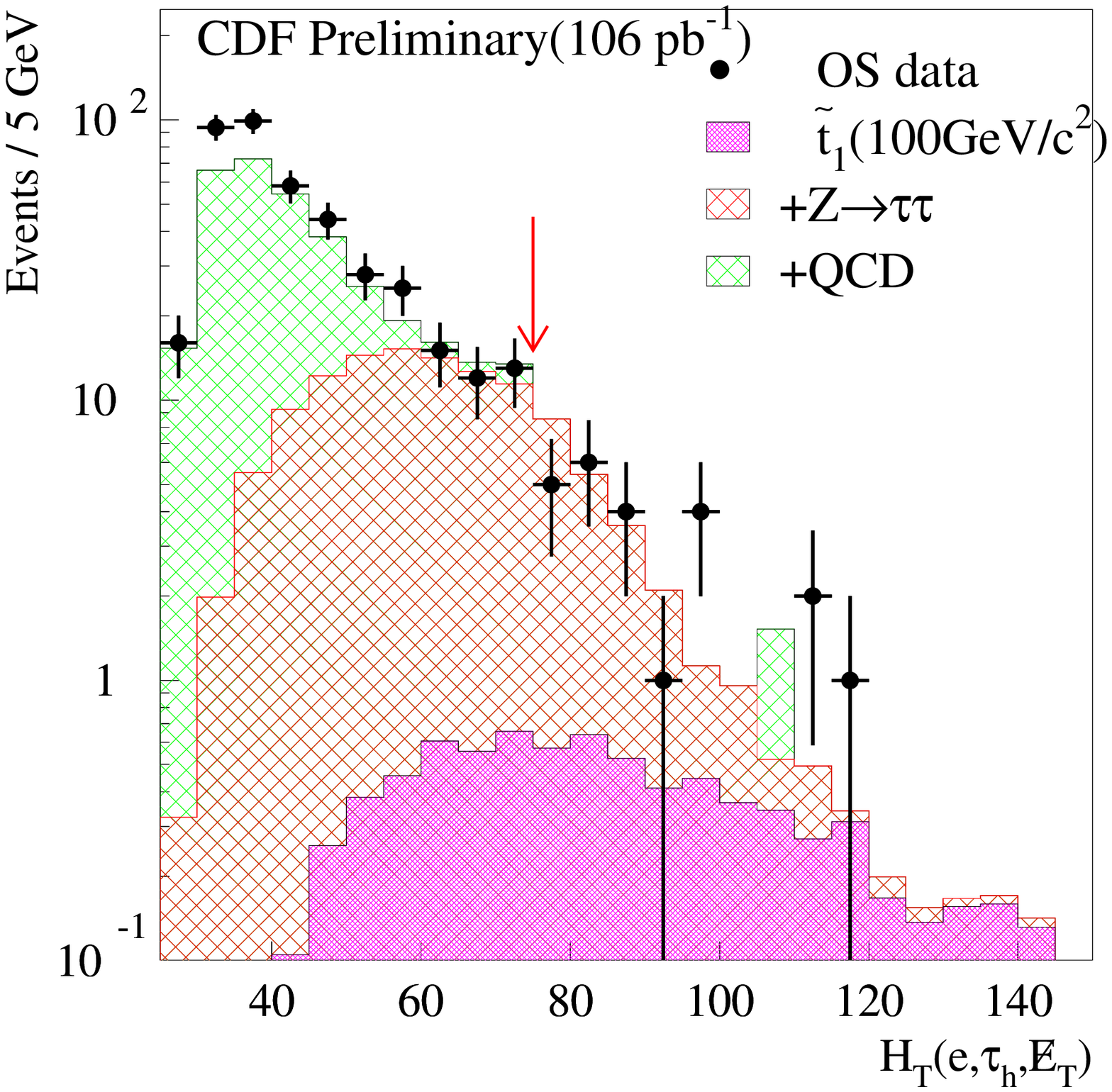,width=\linewidth}
\end{minipage}
\begin{minipage}{0.5\linewidth}
  \centering\epsfig{file=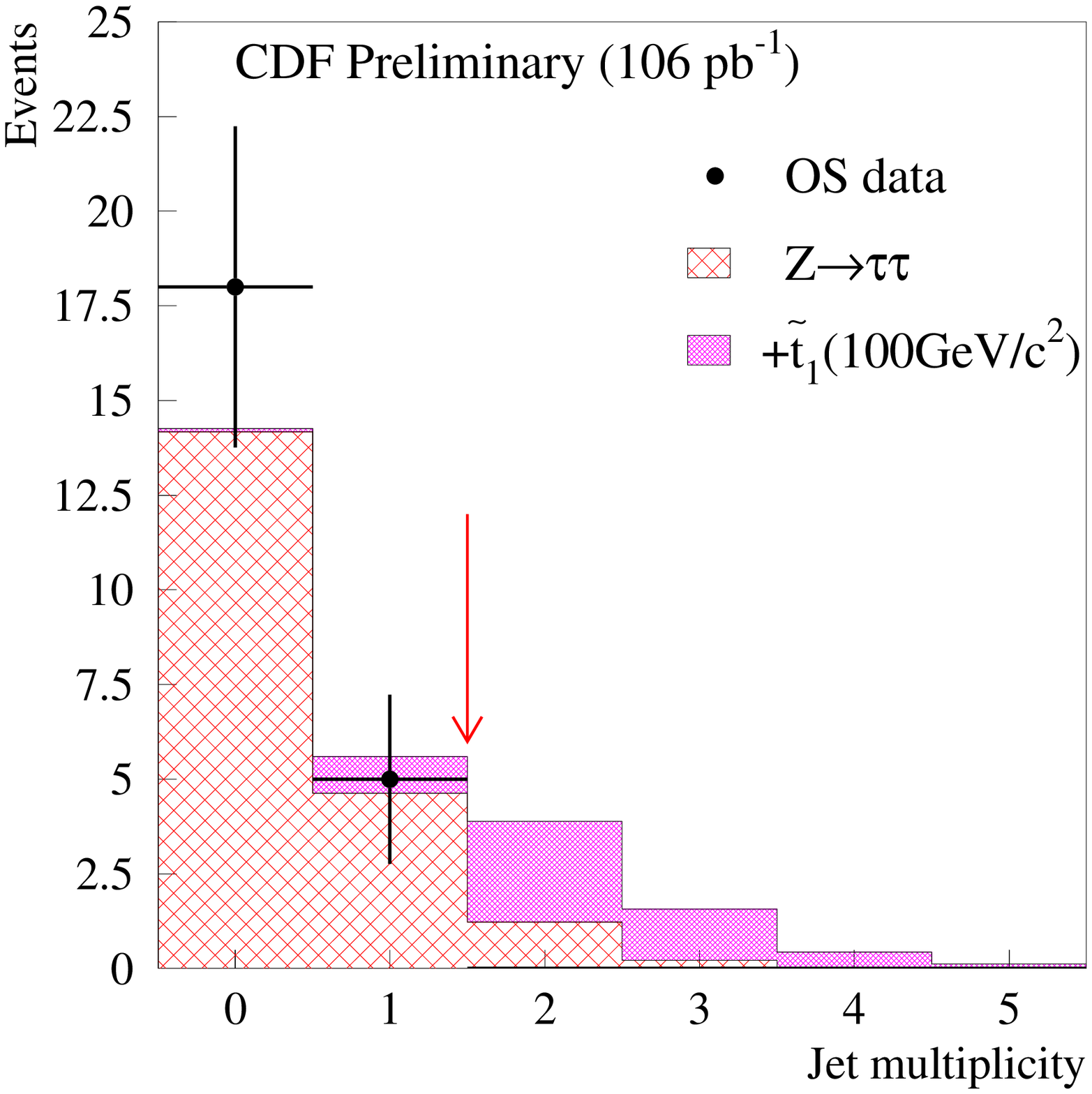,width=\linewidth}
\end{minipage}
  \caption{\it Distribution of $M_{T}({\rm electron, \MET})$ (top left),
  $H_{T}({\rm electron}, \tau_{h}, \mbox{\MET})$ (top right), and
  $N_{jet}$ (bottom) for signal and backgrounds in the electron channel
  of search for RPV m{\sc SUGRA} in decays of stop pairs.}
  \label{fig:rpv_stop_echan}
\end{figure}

The distribution of these variables are shown in
Figure~\ref{fig:rpv_stop_echan}.

\begin{figure}[h!tb] \centering
	\epsfig{file=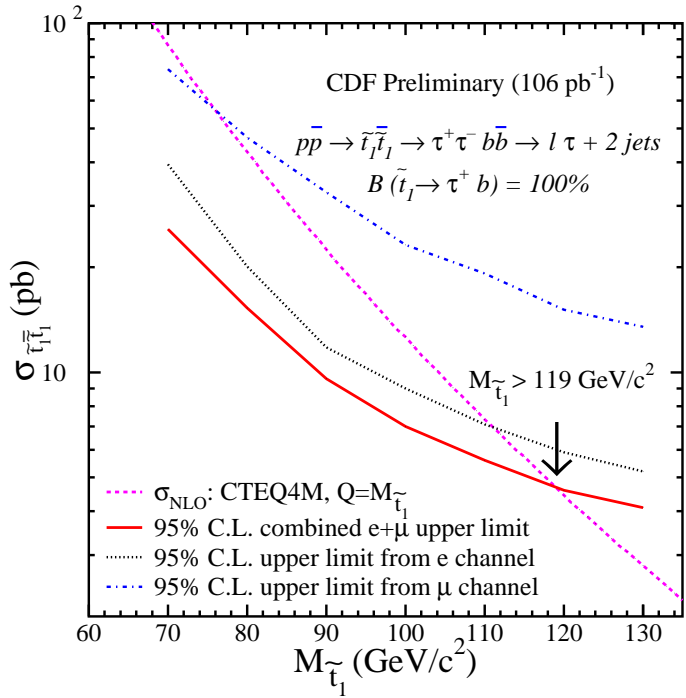, height=10cm}
	\caption{\it Limits in $\sigma_{\tilde{t}\overline{\tilde{t}}}$
	in search for RPV m{|sc SUGRA} in decays of stop pair. We set
	the 95\% C.L. mass limit of stop: $m_{\tilde{t}} > 119 \; {\rm
	GeV}$.}  \label{fig:rpv_stop_combined_limit}
\end{figure}

The results after the cuts are shown in
Table~\ref{tbl:rpv_stop_result}. Since no signal is found in our
data. We set the signal limit in terms of
$\sigma_{\tilde{t}\overline{\tilde{t}}}$, the production cross section
of $\tilde{t}\overline{\tilde{t}}$ as a function of $m_{\tilde{t}}$. The
limits are shown in Figure~\ref{fig:rpv_stop_combined_limit}. From the
figure, we set the 95\% C.L. limit on stop mass: $m_{\tilde{t}} > 119 \;
{\rm GeV}$. The previous limit from ALEPH collaboration is
$m_{\tilde{t}} > 93 \; {\rm GeV}$\cite{rpv_stop_aleph}.

\begin{table} [h!tb]
\centering
\caption{\it Number of expected background events ${\rm N_{bkgd}}$,
observed events ${\rm N_{obs}}$ in data and the expected total signal
efficiency (for $m_{\tilde{t}} = 120 \; {\rm GeV}$) after all the cuts
in search for RPV m{\sc SUGRA} in decays of stop pair are applied.}
\vskip 0.1 in
\begin{tabular}{|c|c|c|c|} \hline
channel & ${\rm N_{bkgd}}$ & ${\rm N_{obs}}$ &
$\varepsilon_{\tilde{t}\overline{\tilde{t}}}$ (\%) \\ \hline \hline
$e$  & $1.92 \pm 0.18$ & 0 & 3.18 \\
$\mu$ & $1.13 \pm 0.13$ & 0 & 1.79 \\
\hline
\end{tabular}
\label{tbl:rpv_stop_result}
\end{table}

\subsection{D\O\ search for resonant slepton in RPV m{\sc SUGRA}}
\begin{figure}[h!tb] \centering
	\epsfig{file=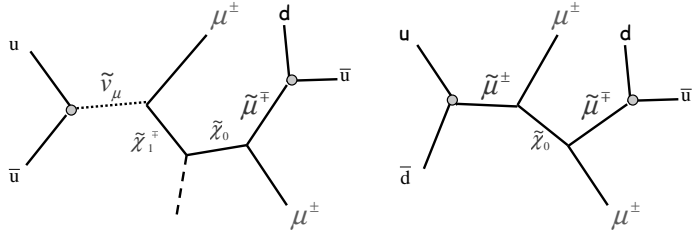,height=10cm,angle=270}
	\caption{\it Feynman diagrams for resonant smuon and muon
	sneutrino production.}
	\label{fig:res_slepton_feynman}
\end{figure}

Resonant sleptons can be produced in the framework of RPV m{\sc SUGRA}. With
the assumption that $\lambda'_{211}$ dominates, the production processes
are shown in Figure~\ref{fig:res_slepton_feynman}. We search for
signatures of resonant slepton in the final states which contain 2 muons
and 2 jets. We apply the following cuts on $94 \; {\rm pb^{-1}}$ of data:

\begin{itemize}
\item $E^{j}_{T} > 20 \; {\rm GeV \; (2 \; jets)}$, $|\eta^{j}| < 2.5$;
\item $P^{\mu}_{T} > 20 \; {\rm Gev/c}$, $|\eta^{\mu_{1}}|, \;
|\eta^{\mu_{2}}| <1.0, \; 1.7$;
\item $H_{T} \; ({\rm all \; jets})> 50 \; {\rm GeV}$;
\item cosmic ray rejection.
\end{itemize}

The major SM background events come from $Z +$~2~jets, $t\overline{t}$
and $WW$ production. After the cuts above, the expected number of
background events and the observed number of data events are listed in
Table~\ref{tbl:res_slepton_events}.

\begin{table} [h!tb]
\centering
\caption{\it Number of expected background events and the number of
observed data events in search for resonant sleptons in the framework of
RPV m{\sc SUGRA} after all the analysis cuts are applied.}
\vskip 0.1 in
\begin{tabular}{|c|c|c|c|c|} \hline
$Z + {\rm 2 jets}$ & $t\overline{t}$ & $WW$ & Total & Observed \\ \hline \hline
4.8 & 0.53 & 0.01 & $5.34 \pm 0.07$ & 5 \\
\hline
\end{tabular}
\label{tbl:res_slepton_events}
\end{table}

\begin{figure}[h!tb] \centering
	\epsfig{file=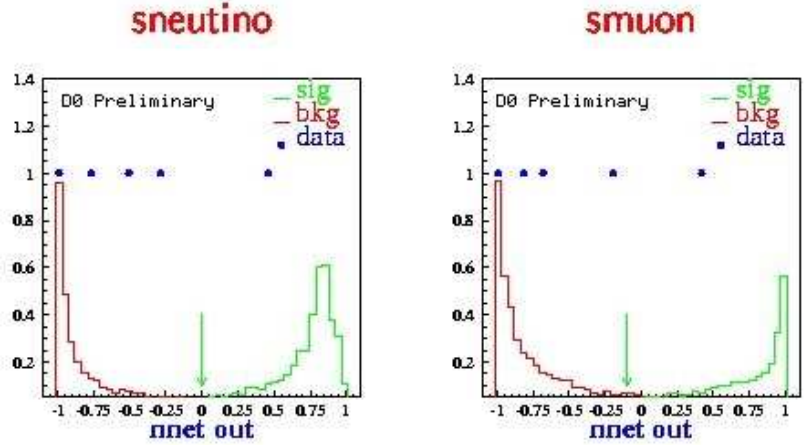, height=8cm}
	\caption{\it Neural Network output for signal (histograms peaked
	near 1), background (histograms peaked at -1), and data
	(points). The plot on the left is for the smuon channel, and the
	plots on the right is for the muon sneutrino channel.}
	\label{fig:res_slepton_nn}
\end{figure}

\begin{figure}[h!tb] \centering
	\epsfig{file=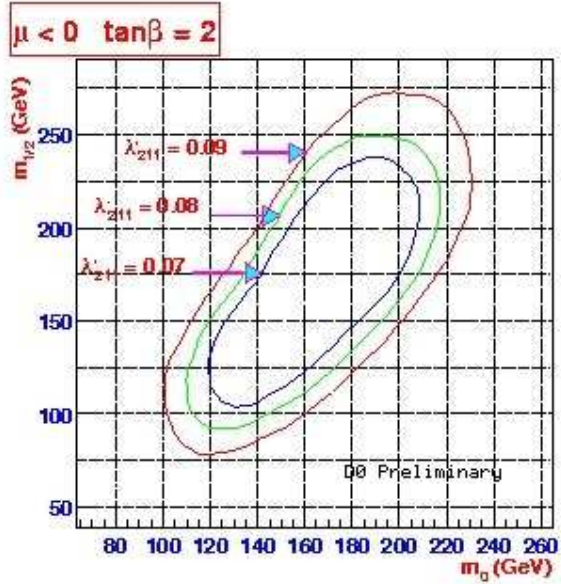, height=8cm}
	\caption{\it 95\% C.L. limit contours in $m_{1/2}-m_{0}$ plane
	obtained in searching for resonant slepton in RPV m{\sc
	SUGRA}. Three $\lambda'_{211}$ coupling constants are shown
	assuming $\mu<0$ and $\tan\beta=2$.}
	\label{fig:res_slepton_limit_tanb2}
\end{figure}

We use Neural Network to further increase the signal significance. The
following seven variables are used in a three-layer Neural Network with
a single node in the output layer\footnote{All Neural Network described
in this paper have this architecture, although the number of
hidden-layer nodes differs from analysis to analysis.}: $E^{j_{1}}_{T} +
E^{j_{2}}_{T}$, $P^{\mu_{1}}_{T} + P^{\mu_{2}}_{T}$,
$M_{\mu_{1},\mu_{2}}$, $\Delta R_{\mu_{1},\mu_{2}}$ (separation of the
two muons in $\eta-\phi$ plane, $\Delta R_{\mu_{1}, j_{nn}}$ ($j_{nn}$
denotes the nearest-neighbor jet), sphericity, and aplanarity. The
Neural Network output are shown in
Figure~\ref{fig:res_slepton_nn}. After applying the cuts indicated by
the arrows in the plots, we expect $1.01 \pm 0.02$ SM background events
and observe 2 in the data. Limit contours in the $m_{1/2}-m_{0}$ plane
for $\mu<2, \; \tan\beta = 2$ is shown in
Figure~\ref{fig:res_slepton_limit_tanb2}. Three coupling
$\lambda'_{211}$ strengths are shown. We are able to exclude $m_{1/2}$
up to 260 GeV for $\lambda'_{211} = 0.09$.

\subsection{D\O\ search for RPV m{\sc SUGRA} in dimuon and four-jets
channel\cite{rpv_2mu4jets}}

In this analysis, we make the following assumptions:
\begin{itemize}
\item {\sc SUSY} particles are pair produced;
\item only one RPV coupling dominates, namely $\lambda'_{222}$;
\item only the LSP, assumed to be the lightest neutralino,
$\widetilde{\chi}^{0}_{1}$, undergoes RPV decay.
\end{itemize}

\begin{figure}[t] \centering
	\epsfig{file=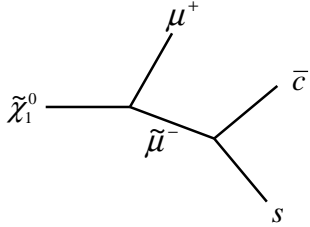,height=7cm,angle=270}
	\caption{\it Feynman diagram of $\widetilde{\chi}^{0}_{1}$ decay
	and the subsequent RPV decay of smuon.}
	\label{fig:rpv_2mu4jets_feynman}
\end{figure}

The relevant Feynman diagram is shown in
Figure~\ref{fig:rpv_2mu4jets_feynman}. We apply the following cuts to
enhance the signal significance in $77.5 \; {\rm pb^{-1}}$ of data:

\begin{itemize}
\item $E^{j}_{T} > 15 \; {\rm GeV}$, $|\eta^{j}| < 2.5$ (4 jets);
\item $P^{\mu_{1}}_{T} > 15 \; {\rm GeV/c}$, $|\eta^{\mu_{1}}| < 1.0$ and
$P^{\mu_{2}}_{T} > 10 \; {\rm GeV/c}$, $|\eta^{\mu_{2}}| < 1.7$,
respectively;
\item $H_{T} \; ({\rm muons \; and \; jets}) > 150 \; {\rm GeV}$;
\item Aplanarity $> 0.03$;
\item $M_{\mu_{1},\mu_{2}} > 5 \; {\rm GeV}$.
\end{itemize}

\begin{figure} [h!tb] \centering
\begin{minipage}{0.5\linewidth}
  \centering\epsfig{file=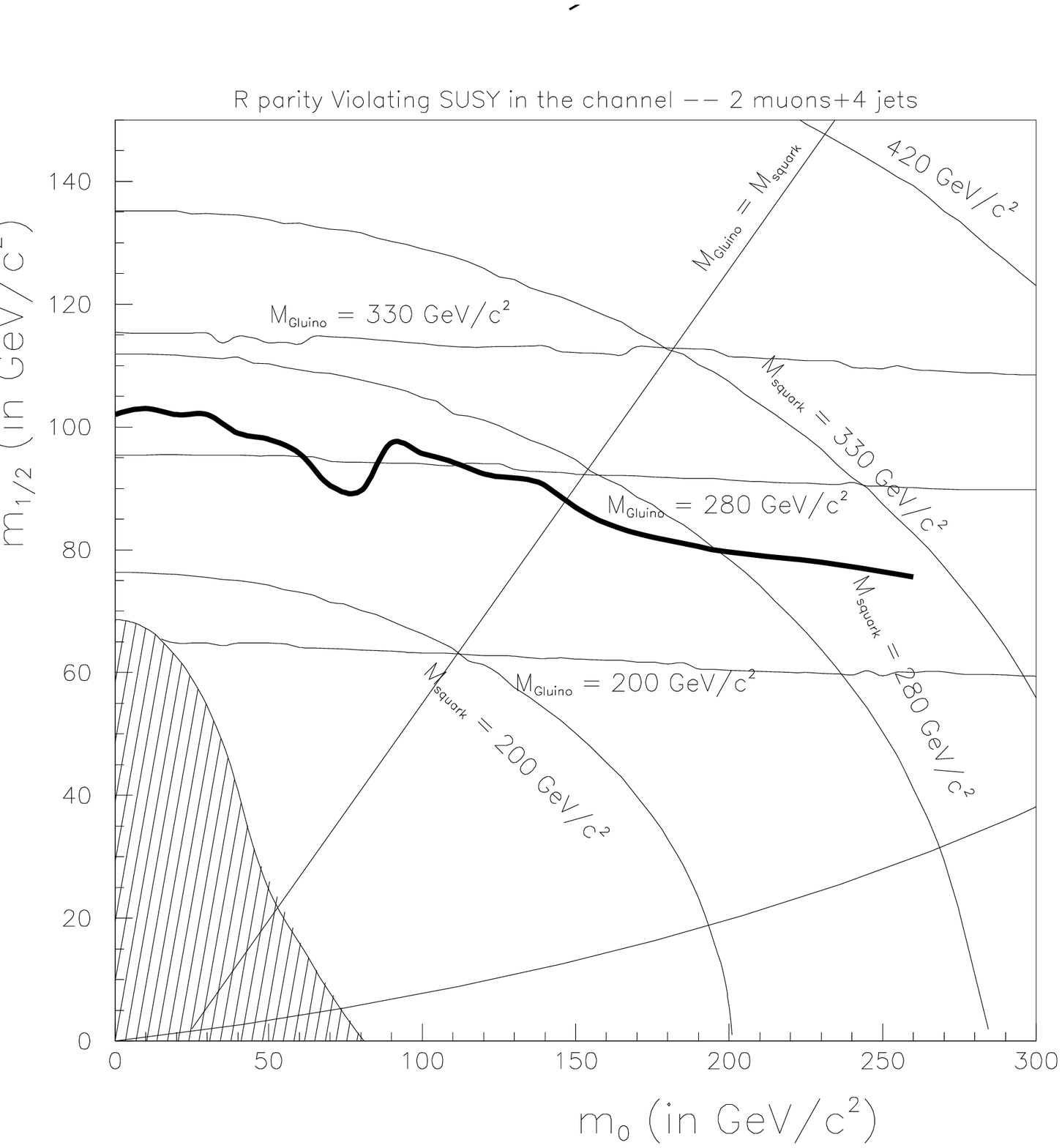,width=\linewidth}
\end{minipage}\hfill
\begin{minipage}{0.5\linewidth}
  \centering\epsfig{file=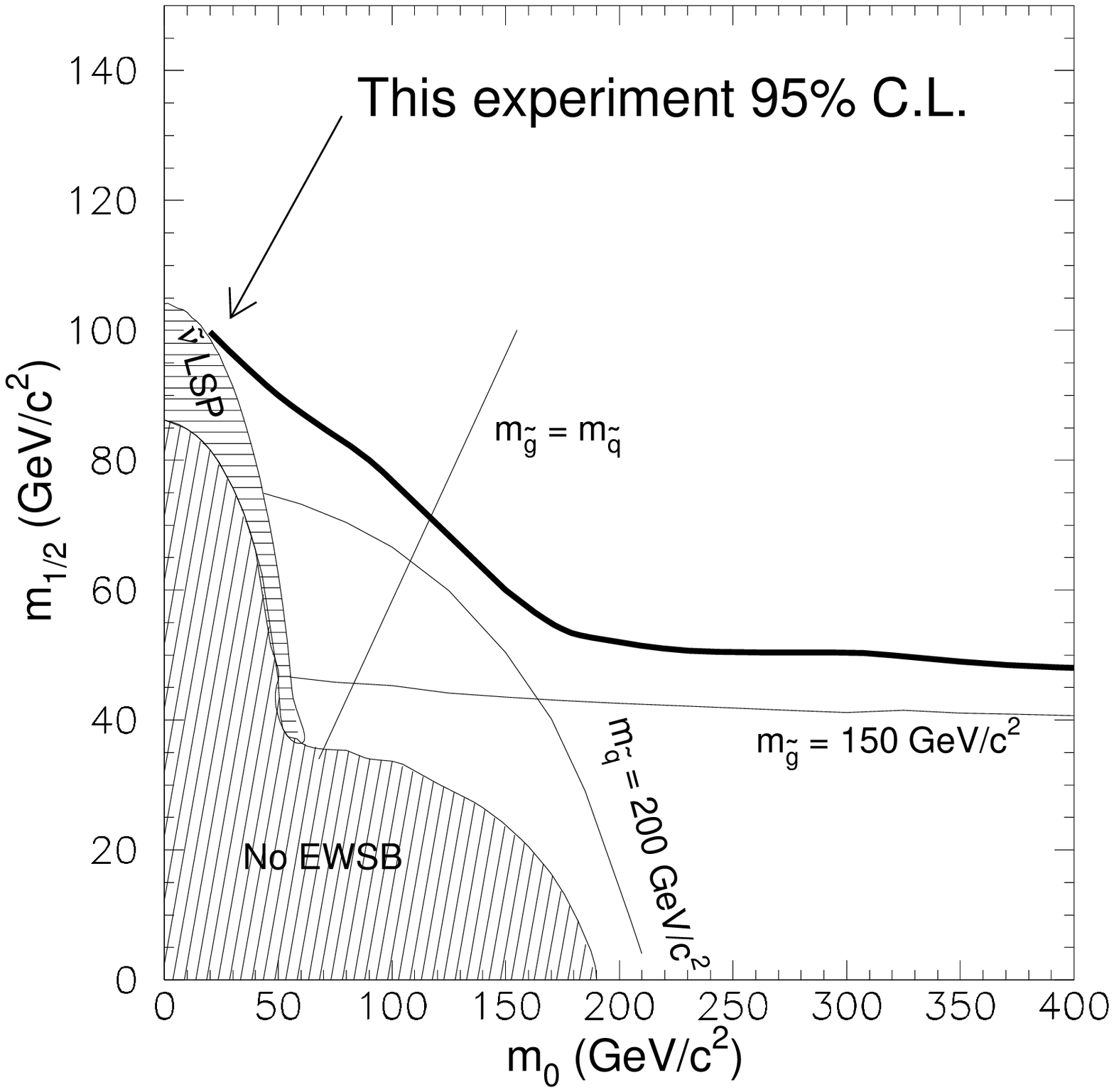,width=\linewidth}
\end{minipage}
  \caption{\it 95\% C.L. limits in $m_{1/2}-m_{0}$ plane for
  $\tan\beta=2$ (plot on the left) and $\tan\beta=6$ (plot on the right)
  in search for RPV m{\sc SUGRA} in dimuon and four-jets channel. We
  assumed that $\mu<0$ and $A_{0}=0$.}
  \label{fig:rpv_2mu4jets_limits}
\end{figure}

The major SM backgrounds are from $Z+ {\rm jets}$ and $t\overline{t}$
processes. The expected number of these events surviving the cuts above
are $0.14 \pm 0.03$ for $Z+ {\rm jets}$ and $0.04 \pm 0.01$ for
$t\overline{t}$, respectively. We observed 0 event in our data. The
typical number of signal events are listed in
Table~\ref{tbl:rpv_2mu4jets_nsig}. Since no signal is seen in our data,
a 95\% C.L. limit contour in $m_{1/2}$-$m_{0}$ plane are set and shown
in Figure~\ref{fig:rpv_2mu4jets_limits}. For the case of $\tan\beta=2$,
we exclude $m_{\tilde{q}} < 240 \; {\rm GeV}$ and $m_{\tilde{g}} < 224
\; {\rm GeV}$.

\begin{table} [h!tb]
\centering
\caption{\it Number of expected signal events after all the selection
cuts for various representative signal parameters in search for RPV
m{\sc SUGRA} in dimuon and four jets channel.}
\vskip 0.1 in
\begin{tabular}{|c|c|c|} \hline
$m_{0}$ & $m_{1/2}$ & ${\rm N_{signal}}$ \\ \hline \hline
80 & 90 & 2.7 \\
190 & 90 & 2.1 \\
260 & 70 & 2.7 \\
400 & 90 & 0.8 \\
\hline
\end{tabular}
\label{tbl:rpv_2mu4jets_nsig}
\end{table}

\subsection{D\O\ search for RPC m{\sc SUGRA} in single electron + $\ge
4$ jets + \MET\ channel\cite{se_msugra}}
\begin{figure}[h!tb] \centering
	\epsfig{file=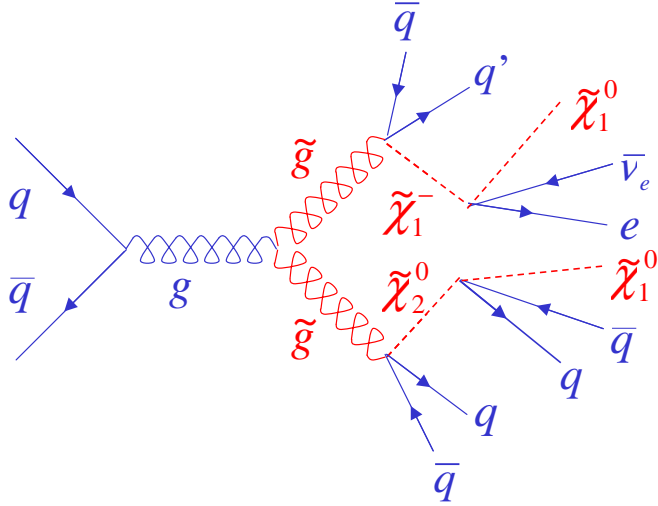,height=10cm,angle=270}
	\caption{\it Feynman diagram for gluino pair production and decay to
        an electron, multijets, and produce \MET. The three-body decays
        are in fact cascade decays in which off-shell particles or
        sparticles are produced. $R_{p}$ is assumed to be conserved.}
	\label{fig:se_feynman}
\end{figure}

D\O\ also conducted a search for $R_{p}$-conserving m{\sc SUGRA} in a
previous unexplored single electron channel. The search is particularly
sensitive to the moderate $m_{0}$ region where charginos and neutralinos
decay mostly into SM $W$ and/or $Z$ bosons which have large branching
fractions to jets. One of the dominant process which produces our
required final state is shown in Figure~\ref{fig:se_feynman}.

The total amount of data used in this search is $92.7 \; {\rm
pb^{-1}}$. The data events are required to pass the following initial
selections:

\begin{itemize}
\item one electron in the good fiducial volume ($|\eta^{e}_{\rm det}| < 1.1$ or
$1.5 < |\eta^{e}_{\rm det}| < 2.5$), and $E^{e}_{T} > 20 \; \rm{GeV}$,
$|\eta^{e}| < 2.0$;
\item no extra electrons in the good fiducial volume with $E^{e}_{T}
> 15 \; \rm{GeV}$;
\item no isolated muons;
\item $E_{T}^{j} > 15 \; \rm{GeV}$, $|\eta^{j}_{\rm det}| < 2.5$ (4 jets);
\item $\mbox{\MET} > 25 \; \rm{GeV}$.
\end{itemize}

The major SM backgrounds and fakes are from $t\overline{t}$, $WW$, $W +
\ge 4 {\rm jets}$, and multijet processes. After the initial selection
cuts above, we observe 72 events in the data and expect $80 \pm 10$
background events. The breakdown of background events in their
respective type is shown in Table~\ref{tbl:se_nevents}.

\begin{table} [h!tb]
\centering
\caption{\it Number of expected background events after initial selection
cuts in search for $R_{p}$-conserving m{\sc SUGRA} in the single
electron + jets + \MET\ channel.}
\vskip 0.1 in
\begin{tabular}{|c|c|c|c|c|} \hline
$t\overline{t}$ & $WW$ & $W W + \ge 4 \; {\rm jets}$ & Multijet & Total \\ \hline \hline
$16.8 \pm 5.2$ & $1.4 \pm 0.3$ & $43.0 \pm 7.6$ & $19.1 \pm 4.7$ & $80 \pm 10$ \\
\hline
\end{tabular}
\label{tbl:se_nevents}
\end{table}

\begin{figure}[h!tb] \centering
	\epsfig{file=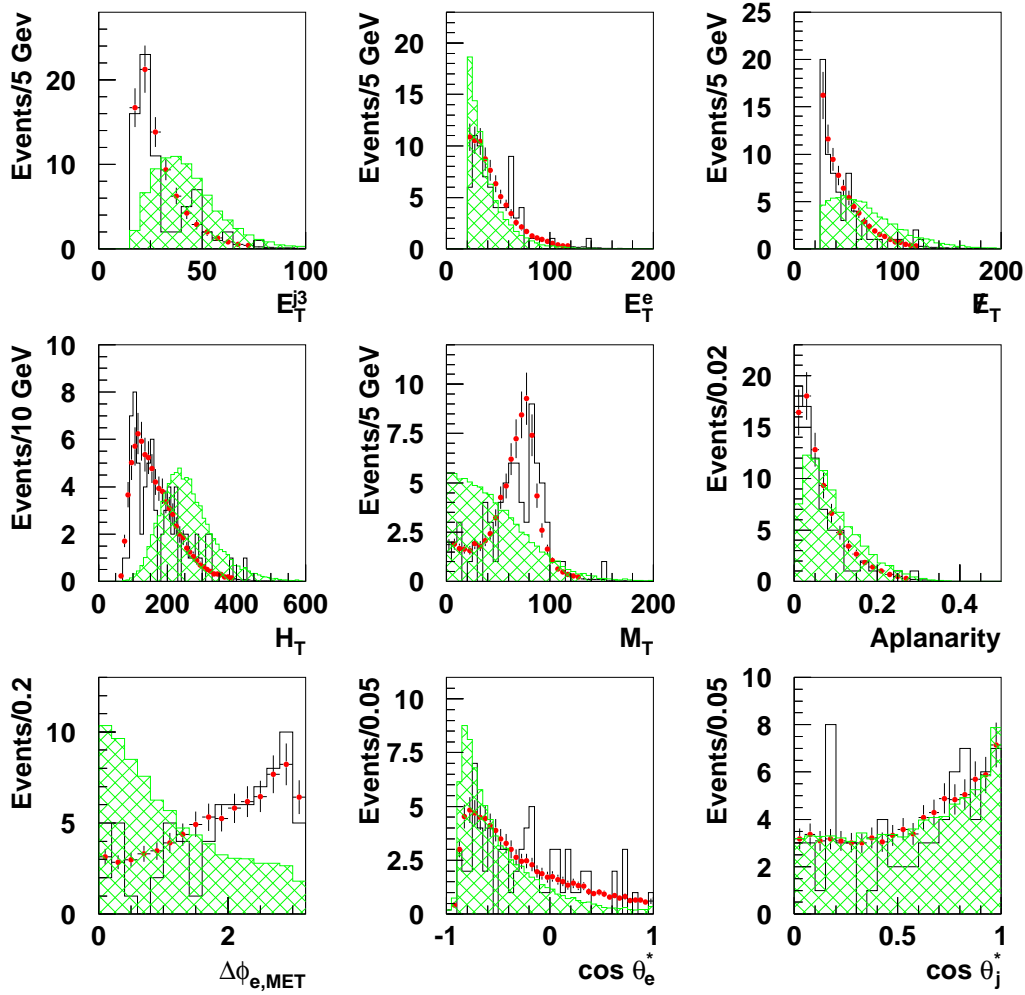,width=\linewidth}
	\caption{\it Distribution of Neural Network input variables for
	signal (hatched histogram), background (open histogram), and
	data (points) in search for $R_{p}$-conserving m{\sc SUGRA} in
	single electron + $\ge 4$ jets + \MET\ channel. The signal and
	background histograms are normalized to the number of expected
	background events ($80 \pm 10$). We observe $72$ events in our
	data.}
	\label{fig:se_nn_var_susy}
\end{figure}

\begin{figure}[h!tb] \centering
	\epsfig{file=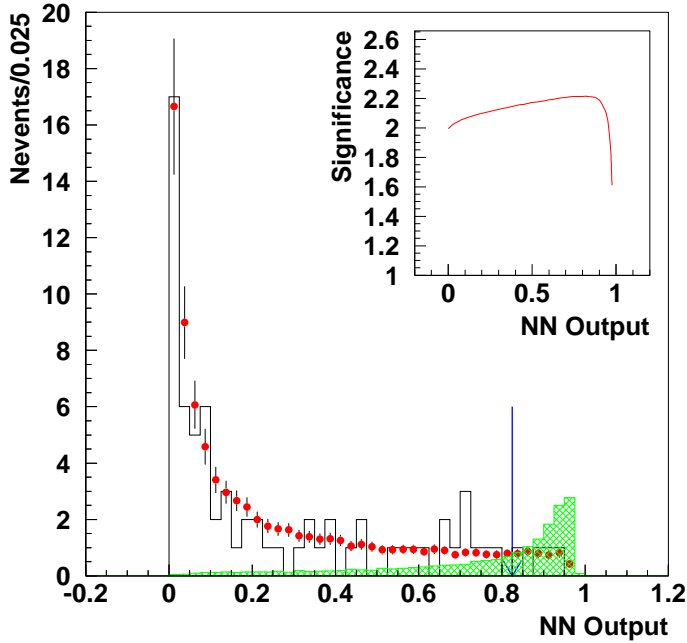,height=10cm}
	\caption{\it Distribution of Neural Network output for signal
	(hatched histogram), background (open histogram), and data
	(points) in search for $R_{p}$-conserving m{\sc SUGRA} in single
	electron + $\ge 4$ jets + \MET\ channel. All histograms are
	normalized to their respective total number of events. The
	insert shows a signal significance as function of Neural Network
	output. The arrow indicates where the signal significance is the
	highest and where we apply our final cut. We retain events to
	the right of the arrow. For this reference signal sample, which
	is generated with $m_{0}=170 \; {\rm GeV}$, $m_{1/2}=58 \; {\rm
	GeV}$, $\tan\beta=3$, $\mu < 0$, and $A_{0} = 0$, $10.4$ signal
	events and $4.4$ background events are expected to survive the
	final cut. We observe $4$ in the data.}
	\label{fig:se_nn_data_susy_bkgd}
\end{figure}

We use Neural Network to enhance the signal significance to set a strong
limit. The input variables to the Neural Network are: $E^{j_{3}}_{T}$,
$E^{e}_{T}$, \MET, $H_{T}$ (all jets), $M_{T}$(e,\MET), aplanarity,
$\Delta\phi_{e,\mbox{\scriptsize \MET}}$, $\cos\theta^{*}_{j}$, and
$\cos\theta^{*}_{j}$, where $\theta^{*}_{j(e)}$ is the polar angle of
the higher-energy jet (electron) from $W$ boson decay in the rest frame
of the parent $W$ boson, relative to the direction of flight of the $W$
boson. This is calculated by fitting the events to the $t\overline{t}$
assumption. The distribution of these variables for signal, background,
and data are shown in Figure~\ref{fig:se_nn_var_susy}. The signal in the
plot is generated by {\sc SPYTHIA}\cite{spythia} with parameters:
$m_{0}=170 \; {\rm GeV}$, $m_{1/2}=58 \; {\rm GeV}$, $\tan\beta=3$, $\mu
< 0$, and $A_{0} = 0$. Both signal and background distributions are
normalized to the total number of expected background events ($80 \pm
10$). The distribution for the Neural Network output is shown in
Figure~\ref{fig:se_nn_data_susy_bkgd}. Shown in the insert of that
figure is a plot of signal significance as a function of Neural Network
output. We apply our final cut on Neural Network output at the highest
signal significance. For the reference signal sample mentioned above, we
cut at NNoutput = 0.825. We expect 10.4 signal events, 4.4 background
events and observe 4 data events after this NNoutput cut.

\begin{figure}[h!tb] \centering
	\epsfig{file=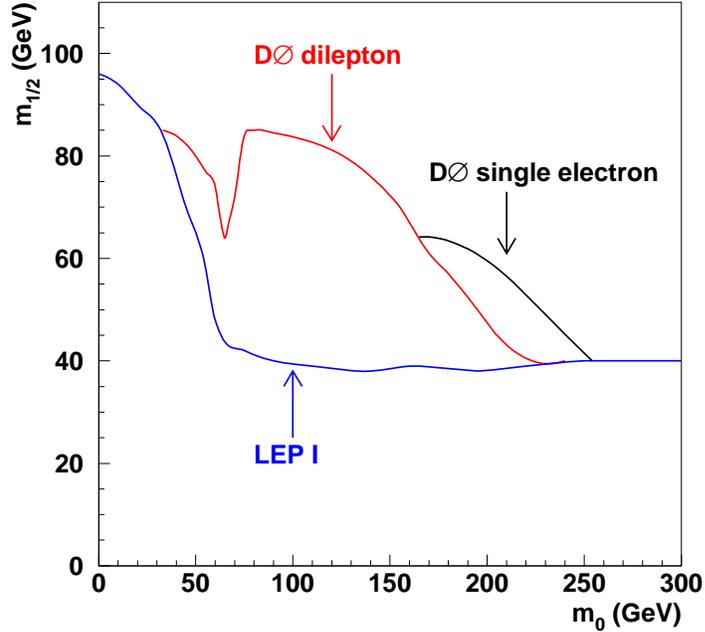,height=10cm} \caption{\it
	95\% limit contour in $m_{1/2}-m_{0}$ plane in search for
	$R_{p}$-conserving m{\sc SUGRA} in single electron + $\ge 4$
	jets + \MET\ channel with $\tan\beta=3$, $\mu < 0$, and $A_{0} =
	0$. Also shown in the plot are limits from LEP I and a previous
	D\O\ search in the dilepton channel.}
	\label{fig:se_tb3_excl}
\end{figure}

Since no signal is observed in our data, a limit is set. The limit
contour is presented in $m_{1/2}-m_{0}$ plane for $\tan\beta=3$, $\mu <
0$, and $A_{0} = 0$ in Figure~\ref{fig:se_tb3_excl}. It extends the
limit set by LEP I and a previous D\O\ search in the dilepton channel in
the moderate $m_{0}$ region. The current best limit is set by
LEP\cite{lepsusy} and the best Tevatron limit is set by the CDF
collaboration\cite{cdf_jets_met_msugra}.

\section{Search for Large Extra Dimension}
A theory of Large Extra Dimensions (LED) has recently been proposed to
solve the hierarchy problem\cite{ADD}. It is argued that gravity lives
in $n$ additional large spatial dimensions while the all the fields of
SM are constrained to a three-dimensional brane which corresponds to our
four space-time dimension. The usually weak gravitational interactions
in the three-brane is in fact strong in the $n$ additional dimensions at
an effective Planck Scale $M_{D}$, which is near the weak scale. Gauss'
Law then relates $M_{D}$ with $n$ and the Planck scale, $M_{pl}$ as:

\begin{equation}
M_{pl} = 8\pi R^{n} M^{n+2}_{D},
\label{eqn:led_theory}
\end{equation}

\noindent where $R$ is the radius of the compactified space in which the
gravitational interaction is strong\footnote{The space is compactified
because $R\approx 10^{32/n-19} \; {\rm meters}$. For $n=2$, R is of the
order of 1mm.}.

While there are many interesting models to solve the hierarchy problem
in the framework of large extra dimension, we will focus on the one
proposed by Arkani-Hamed, Dimopoulos, and Dvali\cite{ADD}, in the
GRW\cite{GRW} convention. One interesting testable aspect of this theory
is that it predicts the existence of Kaluza Klein (KK) tower of massive
gravitons which can interact with the SM fields on the three-brane. We
report two searches for LED:

\begin{itemize}
\item virtual graviton exchange in diphoton production (CDF);
\item direct graviton production in $q\overline{q} \rightarrow gG$ (D\O).
\end{itemize}

\subsection{CDF search for LED in diphoton events}
The differential cross section of diphoton production at the hadron
collider can be written in Eq.~(\ref{eqn:xsec_diphoton}):

\begin{equation}
\frac{d\sigma}{dM_{\gamma\gamma}} =
\frac{d\sigma}{dM_{\gamma\gamma}} \Bigg\vert_{SM} +
\eta\frac{d\sigma}{dM_{\gamma\gamma}} \Bigg\vert_{int} +
\eta^{2}\frac{d\sigma}{dM_{\gamma\gamma}} \Bigg\vert_{KK},
\label{eqn:xsec_diphoton}
\end{equation}

\noindent where $\eta = \frac{\lambda}{M_{D}}$, and $\lambda$ is of
order 1. The three terms in Eq.~(\ref{eqn:xsec_diphoton}) are
contributions from the SM alone, interference of SM and LED, and LED
itself, respectively. Our selection requirements are:

\begin{itemize}
\item $E^{\gamma}_{T} > 22 \; {\rm GeV}$ and $|\eta^{\gamma}_{det}| <
1.0$ for central-central (CC)~\cite{cdfnim} diphoton events;
\item $E^{\gamma_{c}}_{T} > 25 \; {\rm
GeV}$, $|\eta^{\gamma_{c}}_{det}| < 1.0$ and $E^{\gamma_{p}}_{T} > 22 \;
{\rm GeV}$, $1.1 < |\eta^{\gamma_{p}}_{det}| < 2.4$ for central-plug
(CP)\cite{cdfnim} diphoton events.
\end{itemize}

\begin{figure}[h!tb] \centering
	\epsfig{file=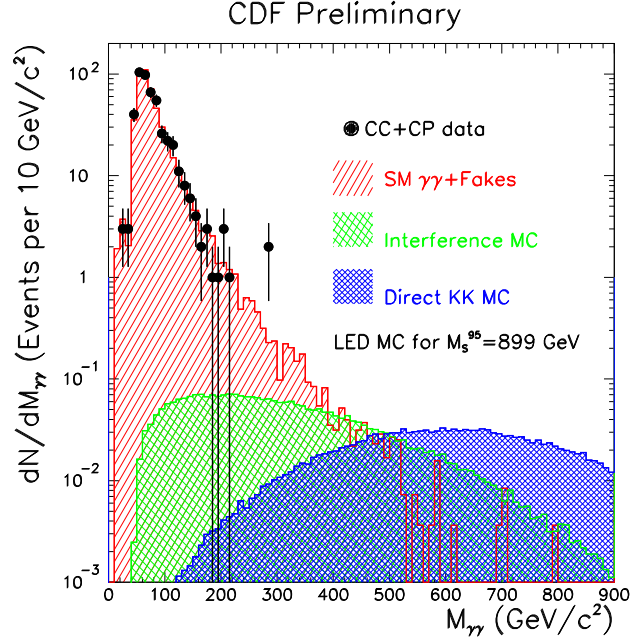,height=10cm}
	\caption{\it Invariant mass spectrum of diphoton events for
	SM+Fakes, direct LED contribution, the LED/SM interference, and
	data in search for LED in the photon + \MET\ channel.}
	\label{fig:led_diphoton_limit}
\end{figure}

After these selections we expect that in the $100 \; {\rm pb^{-1}}$ of
data we use for this analysis, there are $96 \pm 31$ SM events
and $184 \pm 63$ fake events in CC and $76 \pm 31$ SM events and $132
\pm 28$ fake events in EC. We observe 287 and 192 events, respectively. Since
no excess is seen in our data, we perform a log-likelihood fit on the
diphoton invariant mass spectrum to Eq.~(\ref{eqn:xsec_diphoton}) to
extract the 95\% C.L. limit on $M_{D}$. The diphoton invariant mass
spectra of data, background, and signal are shown in
Figure~\ref{fig:led_diphoton_limit}. From the fit, we obtain $M_{D} >
1.01 \; {\rm TeV}$. Note that D\O\ performed a similar search which
included dielectron events in the data set ($127 \; {\rm pb^{-1}}$). In
addition to invariant mass, the polar angle of the electron or photon in
the rest frame of the dielectron or diphoton system was also used as a
parameter in the fit. The extracted limit is $M_{D} > 1.21 \; {\rm
TeV}$\cite{d0_led_2e_2g}.

\subsection{D\O\ search for direct graviton production in LED in the
monojet channel}

Graviton in LED can also be produced at the Tevatron directly through
$q\overline{q} \rightarrow gG$ and $gg \rightarrow gG$. The distinct
signature in the detector is a single jet (monojet) with large missing
transverse energy.

\begin{figure}[h!tb] \centering
	\epsfig{file=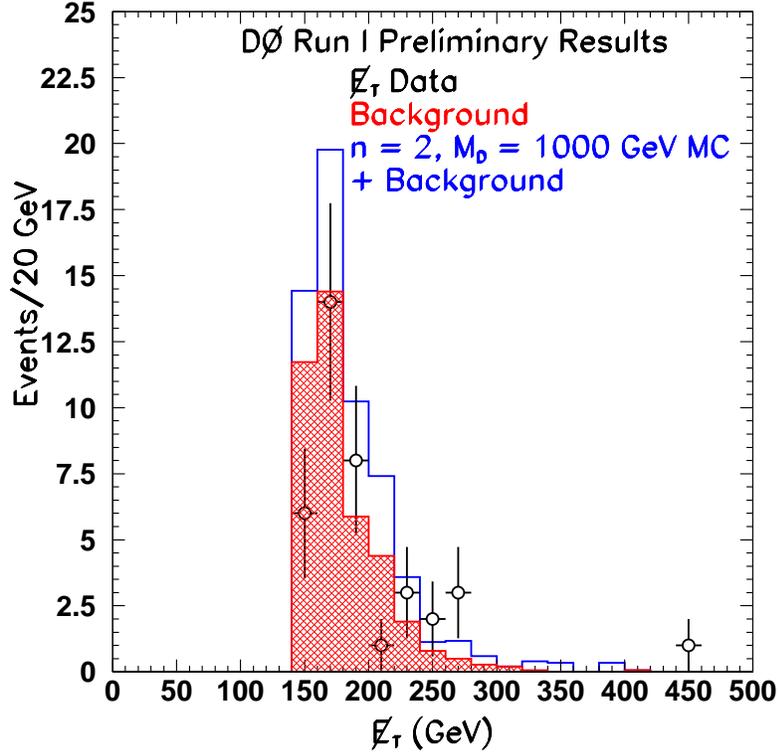,height=10cm}
	\caption{\it \MET\ spectrum for background (hatched histogram),
	expected signal for $n=2$ and $M_{D}=1000 \; {\rm GeV}$ (stacked
	open histogram), and data (points) in search for LED in the
	monojet channel. The data and background agree very well.}
	\label{fig:led_jet_met_data_bkgd_signal}
\end{figure}

\begin{figure}[h!tb] \centering
	\epsfig{file=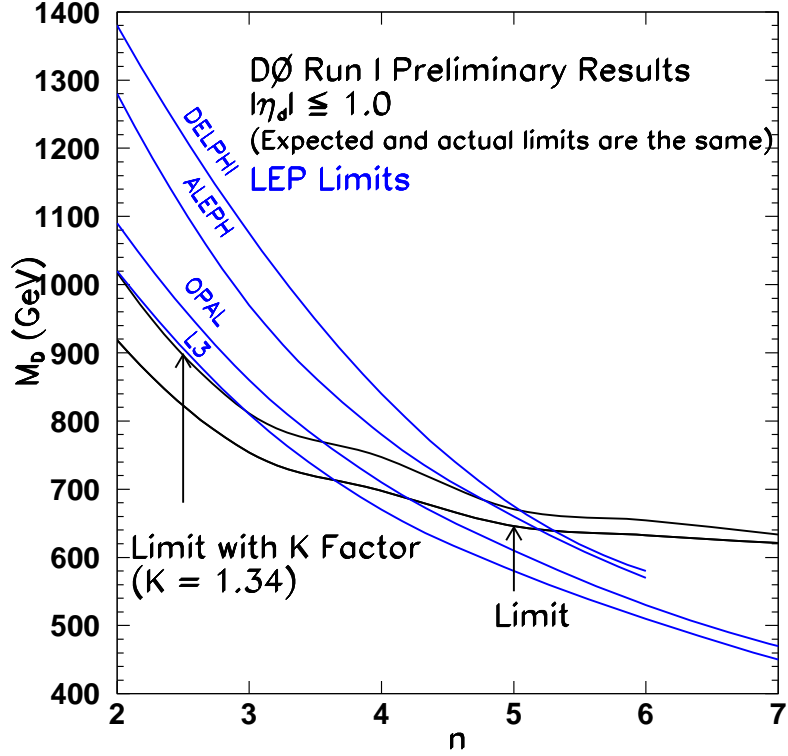,height=10cm}
	\caption{\it 95\% C.L. lower limits on $M_{D}$ in the D\O\
	search for LED in the monojet channel. Limits with and without
	k-factor ($k = 1.34$) are plotted. Also shown is the limits
	obtained by the LEP experiments.}
	\label{fig:led_jet_met_limit}
\end{figure}

In this D\O\ search, the signal events are generated using a subroutine
written by Lykken and Matchev\cite{lykken_matchev} as the external
process to PYTHIA\cite{pythia}. To reduce the backgrounds, which are
dominated by $Z \rightarrow \nu\nu$ and $W (l\nu)$ fake events, D\O\
require that $E^{j}_{T} > 150 \; {\rm GeV}$, $|\eta^{j}_{det}| < 1.0$,
and $\mbox{\MET} > 150 \; {\rm GeV}$. We also reject events with
isolated muons or with $E^{j_{2}}_{T} > 50 \; {\rm GeV}$, where $j_{2}$
denotes the second leading jet in $E_{T}$ in the event. The total amount
of data used in this analysis is $78.8 \; {\rm pb^{-1}}$. After applying
the cuts mentioned above, we observe 38 event and expect $30.2 \pm 4.0$
$WZ$ SM events and $7.8 \pm 7.1$ multijet and cosmic fake events. The
\MET\ spectrum after the cuts is shown in
Figure~\ref{fig:led_jet_met_data_bkgd_signal}. The expected total
background agrees with data very well. We thus set the 95\% C.L. limits
on $M_{D}$ as shown in Figure~\ref{fig:led_jet_met_limit}. Also plotted
in the figure are limits obtained by LEP experiments. We note that at
higher extra dimensions $n \ge 5$, because the large center-of-mass
energy the Tevatron can reach, D\O\ limit extends those obtained by the
LEP experiments. The numeric value of the limit at each extra dimension
is also listed in Table~\ref{tbl:led_jet_met_limit}.

\begin{table} [h!tb]
\centering
\caption{\it 95\% C.L lower limits on the effective Planck scale $M_{D}$
as a function of number of extra dimension $n$ obtained by D\O\ search
for LED in monojet channel. No k-factor is applied.}
\vskip 0.1 in
\begin{tabular}{|c|c|c|c|c|c|c|} \hline
$n$ & 2 & 3 & 4 & 5 & 6 & 7 \\ \hline \hline
$M_{D}$ (TeV) & 0.89 & 0.73 & 0.68 & 0.64 & 0.63 & 0.62 \\
\hline
\end{tabular}
\label{tbl:led_jet_met_limit}
\end{table}

\section{Search for leptoquarks in jets + \MET\ channel\cite{lq_jets_met}}
Because of the flavor symmetry between the lepton and quark sectors of
the SM, theories have been developed to explore the possible direct
couplings between the leptons and the quarks through new
particles. Leptoquarks (LQ) are one of these postulated particles. They
are either scalar\cite{lq_scalar} or vector\cite{lq_vector} bosons which
carry both color and fractional electric charge. At the Tevatron, LQ can
be pair produced through strong interaction: $p\overline{p} \rightarrow
LQ\overline{LQ} + X$.

Limits from flavor-changing neutral currents imply that leptoquarks of
low mass ${\cal O}$(TeV) couple only within a single
generation\cite{lq_fcnc_constraint}. Leptoquark pair can decay into
$l^{\pm}l^{\mp}q\overline{q}$, $l^{\pm}\nu q\overline{q}$, and
$\nu\overline{\nu}q\overline{q}$ final states. Both CDF and D\O\ have
performed search in the dilepton and single lepton channels. The
analysis reported here is based on the $\nu\overline{\nu}q\overline{q}$
final state. It is sensitive to leptoquarks of all three
generations. Searches for both scalar and vector leptoquarks are
performed. In the case of vector leptoquark, we consider specific cases
of couplings resulting in the minimal cross section ($\sigma_{\rm
min}$), Minimal Vector coupling (MV), and Yang-Mills coupling
(YM)\cite{lq_vector_couplings}.

The total amount of data used in this analysis is $85.2 \; {\rm
pb^{-1}}$. The following initial cuts are applied:

\begin{itemize}
\item $E^{j_{1}}_{T} > 50 \; {\rm GeV}$, $E^{j_{2}}_T > 40 \; {\rm
GeV}$, and at least one of the jets is in $|\eta|<1.0$;
\item $\mbox{\MET} > 40 \; {\rm GeV}$;
\item $\Delta\phi({\rm all \; jets, \MET}) > 30^{\circ}$,
$\Delta\phi(j_{2}, \mbox{\MET}) > 60^{\circ}$.
\end{itemize}

Events with isolated muons are rejected. After the initial cuts, we are
left with 231 events in the data. The major SM background and fakes come
from multijet, $W +$ jets, $Z +$ jets, and $t\overline{t}$ processes. We
estimate that there are $242 \pm 19 ({\rm stat}) ^{+29}_{-10} ({\rm
sys})$ of such events in the data after the initial cuts. The data and
background estimation agree very well.

\begin{figure} [h!tb] \centering
\begin{minipage}{0.5\linewidth}
  \centering\epsfig{file=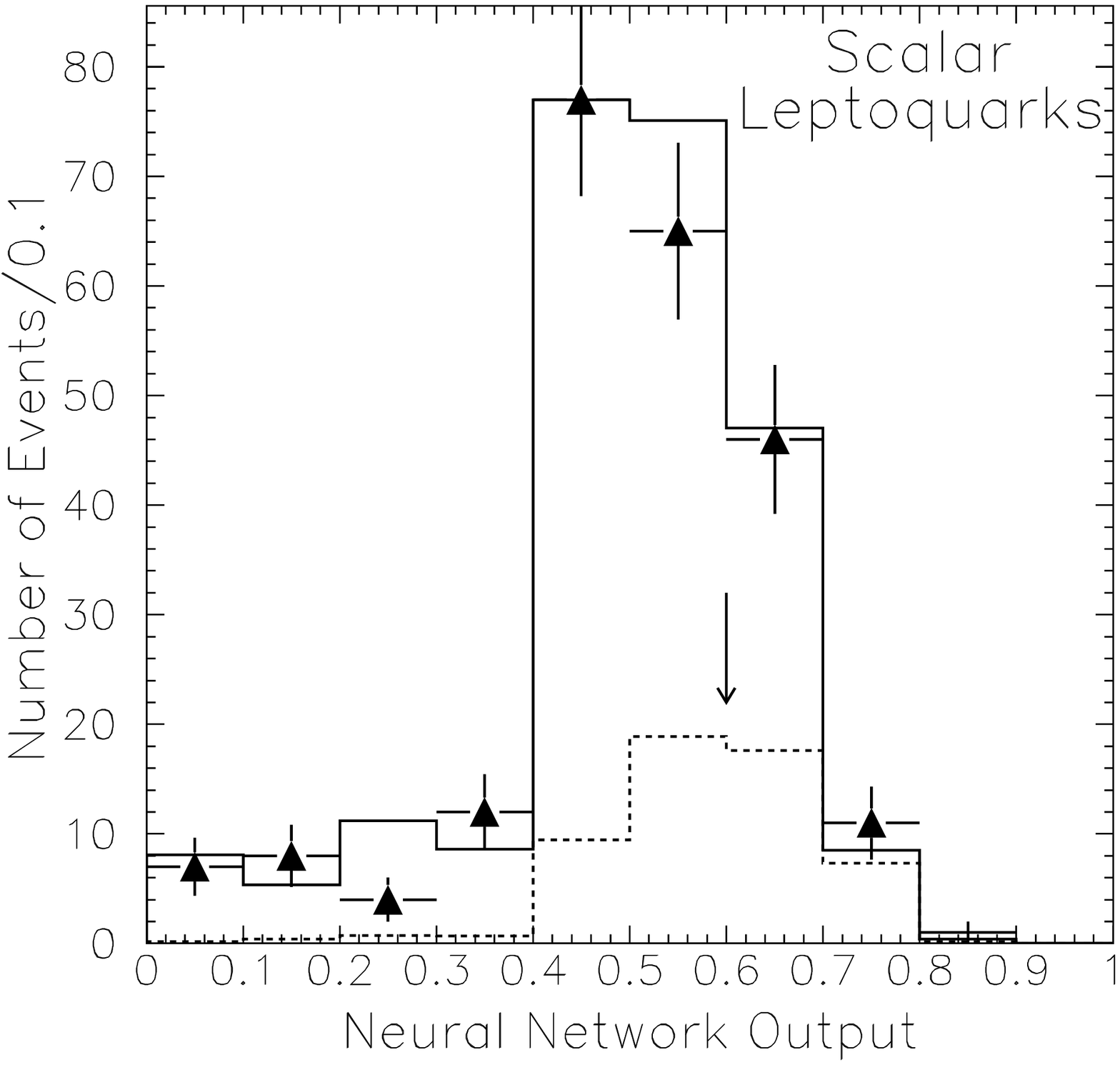,width=\linewidth}
\end{minipage}\hfill
\begin{minipage}{0.5\linewidth}
  \centering\epsfig{file=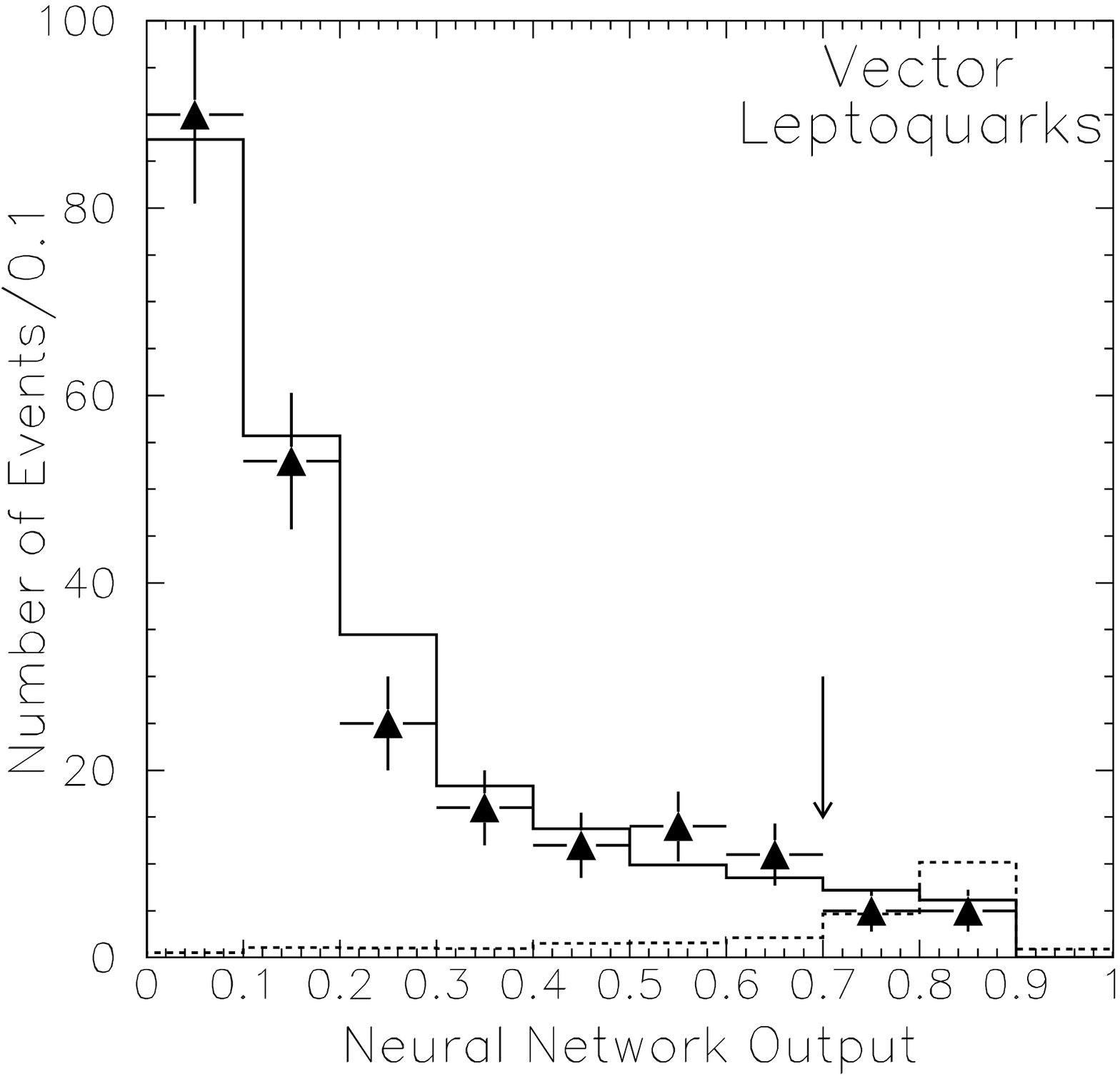,width=\linewidth
}
\end{minipage}
  \caption{\it Neural Network output for scalar (plot on the left) and
  vector (plot on the right) leptoquark signals. The points are data and
  the histograms are the expected background. We apply the final cut at
  the Neural Network output values indicated as arrows in the plots to
  obtain a strong limit.}
  \label{fig:lq_nn}
\end{figure}

\begin{figure} [h!tb] \centering
\begin{minipage}{0.5\linewidth}
  \centering\epsfig{file=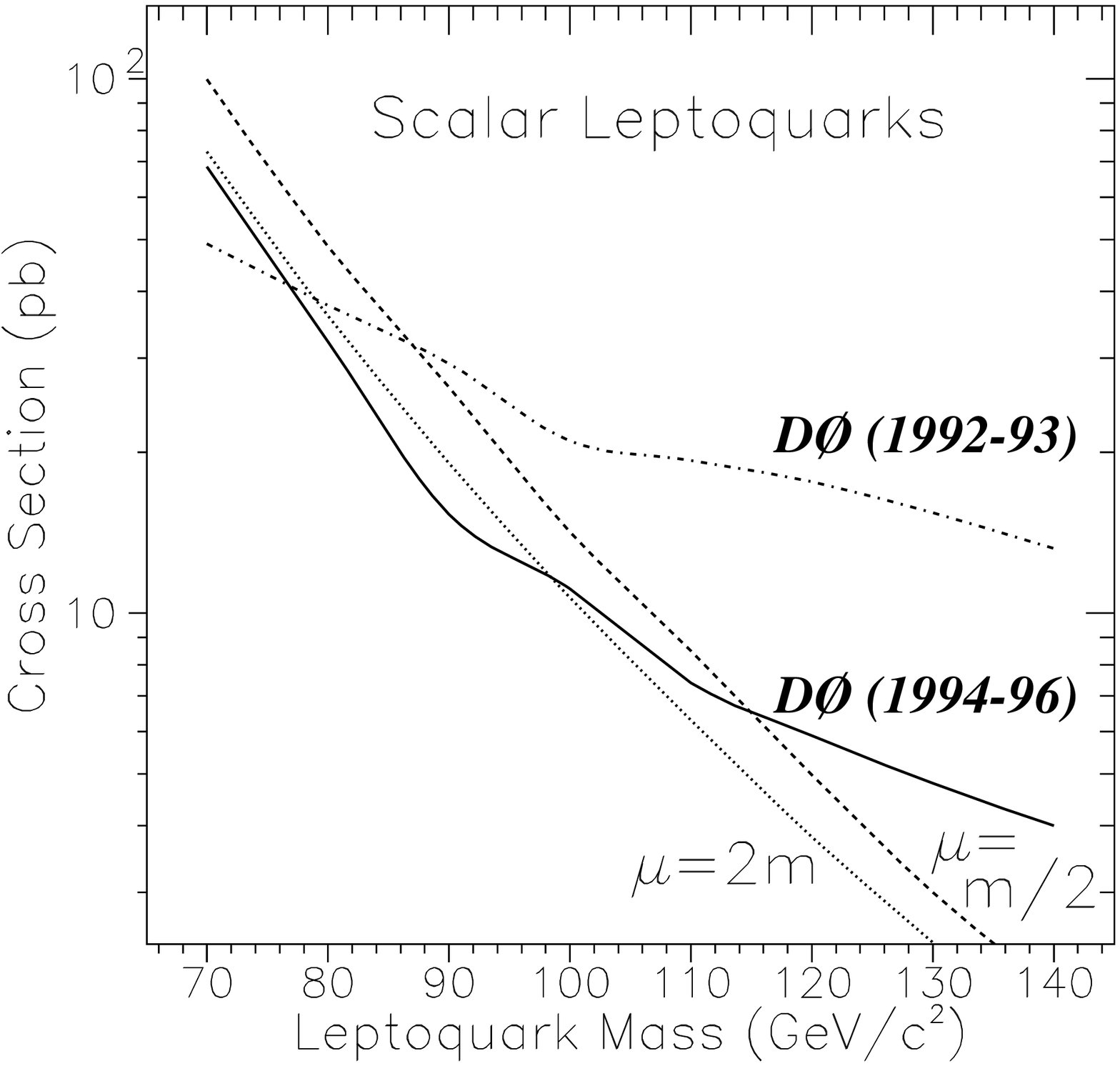,width=\linewidth}
\end{minipage}\hfill
\begin{minipage}{0.5\linewidth}
  \centering\epsfig{file=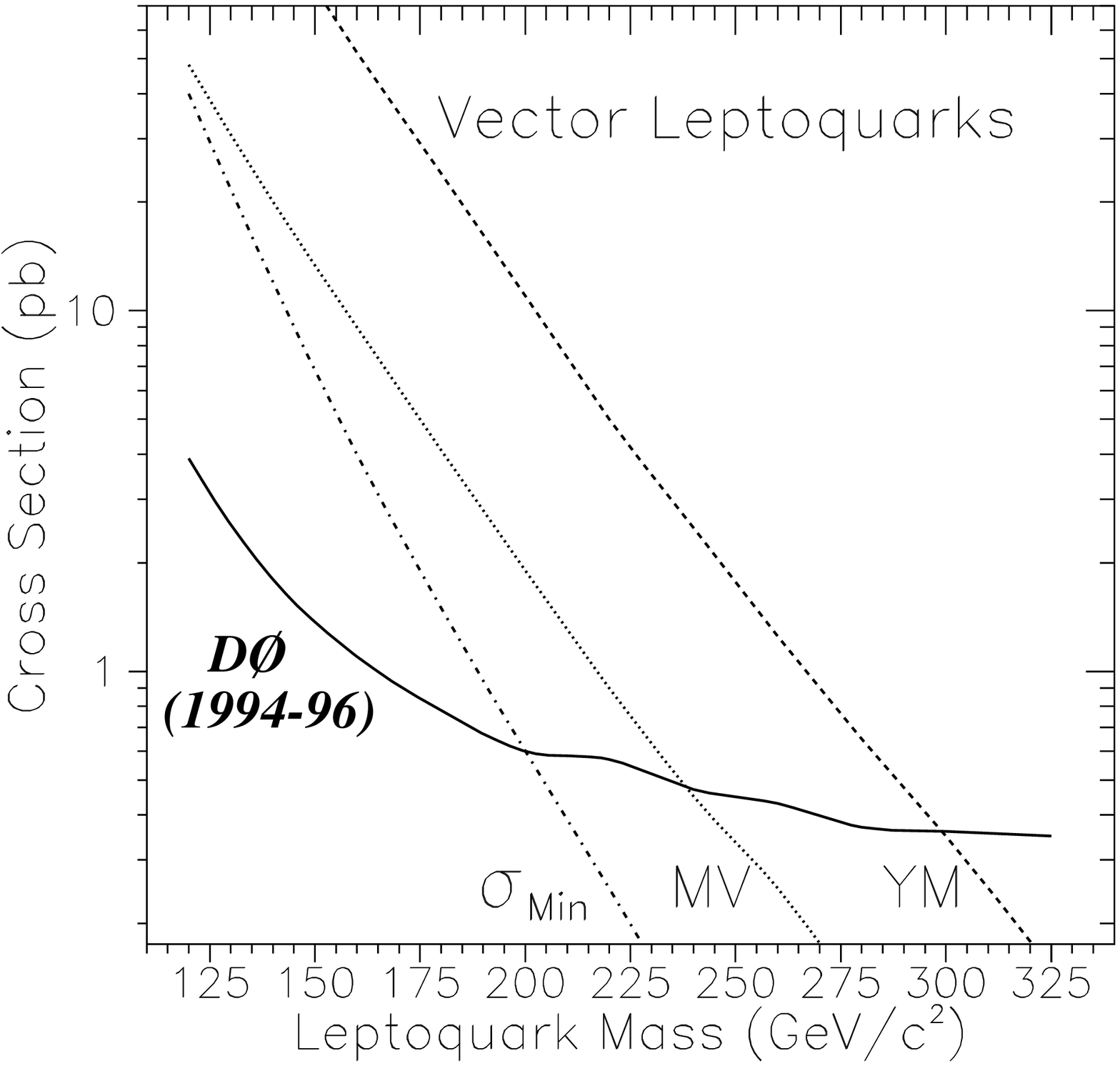,width=\linewidth}
\end{minipage}
  \caption{\it 95\% C.L. limit on leptoquark production cross section as
  a function of leptoquark mass for the scalar (plot on the left) and
  vector (plot on the right) leptoquarks. The dotted curves are
  theoretical calculations and the 95\% confidence limit set by a
  previous D\O\ search~\cite{d0_lq_run1a} using one tenth the amount of
  the data as what are in this analysis.}
  \label{fig:lq_limit}
\end{figure}

\begin{figure} [h!tb] \centering
\begin{minipage}{0.5\linewidth}
  \centering\epsfig{file=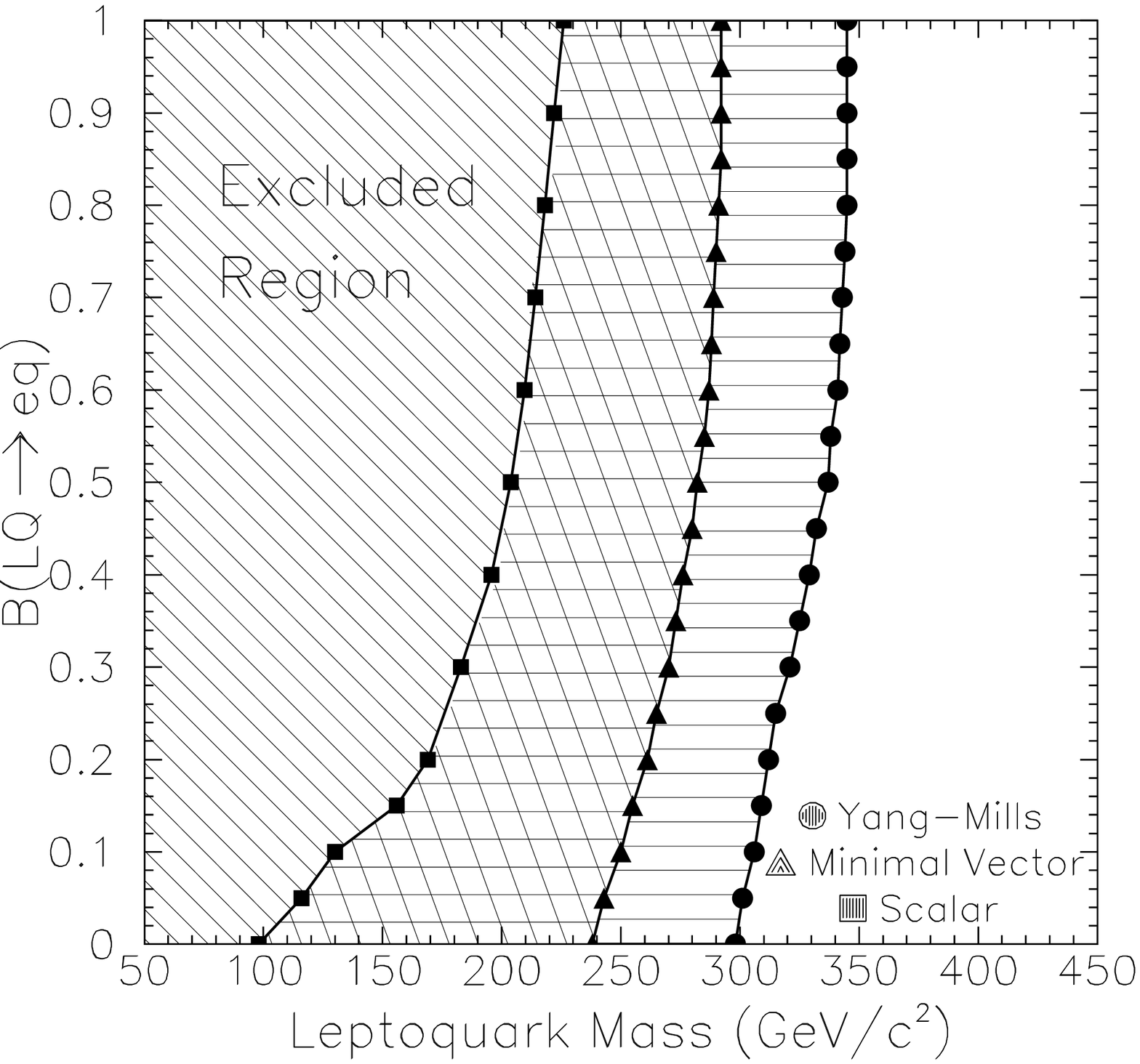,width=\linewidth}
\end{minipage}\hfill
\begin{minipage}{0.5\linewidth}
  \centering\epsfig{file=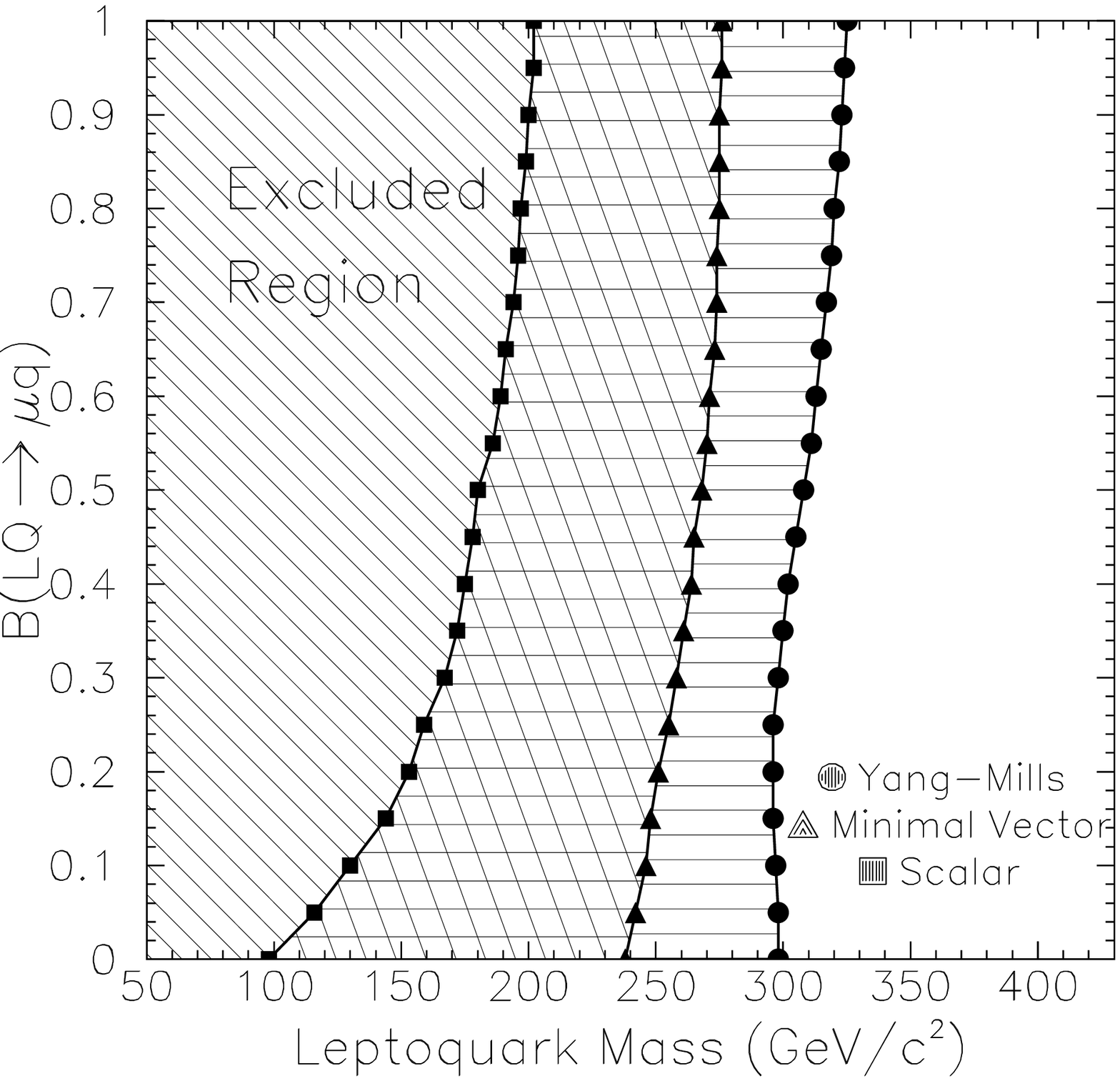,width=\linewidth}
\end{minipage}
  \caption{\it Combined 95\% C.L. limit on scalar (plot on the left) and
  vector (plot on the right) leptoquark mass as a function of leptoquark
  branch fraction into lepton and quark: BR(${\rm LQ} \rightarrow
  l^{\pm}q$). The results combined are those from this analysis, and
  those from previous D\O\ searches for the first and second generation
  leptoquarks.}
  \label{fig:lq_combined_limit}
\end{figure}

In order to set a strong limit, we use Neural Network to enhance the
signal significance. The input variables are \MET\ and $\Delta\phi
(j_{1}, j_{2})$ for the scalar leptoquark signal, and \MET\ and
$E^{j_{2}}_{T}$ for the vector leptoquark signal. The distribution of
the Neural Network output is shown in Figure~\ref{fig:lq_nn}. We
calculate a signal significance as a function of Neural Network
output. The values which correspond to the maximal signal significance
are indicated as arrows in Figure~\ref{fig:lq_nn}. After removing events
to the left of the arrows, we are left with $56 \pm 3$ ($13 \pm 3$)
expected background events and 58 (10) observed data events in the case
of scalar (vector) leptoquark signal. The resulting 95\% C.L. limit
contours are shown in Figure~\ref{fig:lq_limit}. From the limit contours
we obtain the 95\% C.L. limit on the leptoquark masses: $M_{SLQ} > 98 \;
{\rm GeV}$, and $M_{VLQ} > 200, 238, 298 \; {\rm GeV}$ for
$\sigma_{min}$, MV, and YM coupling, respectively. The results of this
analysis are also combined with those obtained in the previous first and
second generation leptoquark searches by D\O\cite{d0_lq_1gen,
d0_lq_2gen}. The resulting mass limits as a function of BR(${\rm LQ}
\rightarrow l^{\pm}q$) are shown in
Figure~\ref{fig:lq_combined_limit}. The gaps at small BR in the previous
analyses are filled as a result of this investigation.

\section{Model Independent Searches}
During the past year, both CDF and D\O\ performed searches that are
mostly based on final states of data rather than specific models that
result in those final states. Understanding the final states makes the
search of models a much more straightforward process.

\subsection{CDF photon + \MET\ analysis\cite{gamma_met}}
Photon + \MET\ is an interesting signal for many new physics. In this
analysis, we first prepare a set of photon + \MET\ data sample with all
the contributions from SM background and fakes understood. We then look
at models which may result in the photon + \MET\ final state. We apply
the following selection cuts to reduce the data to a reasonable size:

\begin{itemize}
\item $E^{\gamma}_{T} > 55 \; {\rm GeV}$ and $|\eta^{\gamma}_{det}|<1.0$;
\item $\mbox{\MET} > 45 \; {\rm GeV}$;
\item no jets with $E^{j}_{T} > 15 \; {\rm GeV}$ or tracks with $P_{T} >
5 \; {\rm GeV/c}$.
\end{itemize}

We observe 11 events and expect 11 background events in $87 \; {\rm
pb^{-1}}$ of data. The background sources and their expected number of
events surviving the cuts are listed in Table~\ref{tbl:photon_met_bkgd}.

\begin{table} [h!tb]
\centering
\caption{\it SM background sources and their expected number
of events in $87 \; {\rm pb^{-1}}$ of photon + \MET\ data.}
\vskip 0.1 in
\begin{tabular}{|c|c|} \hline
Background source & Events \\ \hline \hline
cosmic rays & $6.3 \pm 2.0$ \\
$Z\gamma \rightarrow \nu\nu\gamma$ & $3.2 \pm 1.0$ \\
$W \rightarrow e\nu$ & $0.9 \pm 0.1$ \\
prompt diphoton & $0.4 \pm 0.1$ \\
$W\gamma$ & $0.3 \pm 0.1$ \\
Total & $11.0 \pm 2.2$ \\
\hline
\end{tabular}
\label{tbl:photon_met_bkgd}
\end{table}

\begin{figure}[h!tb] \centering
	\epsfig{file=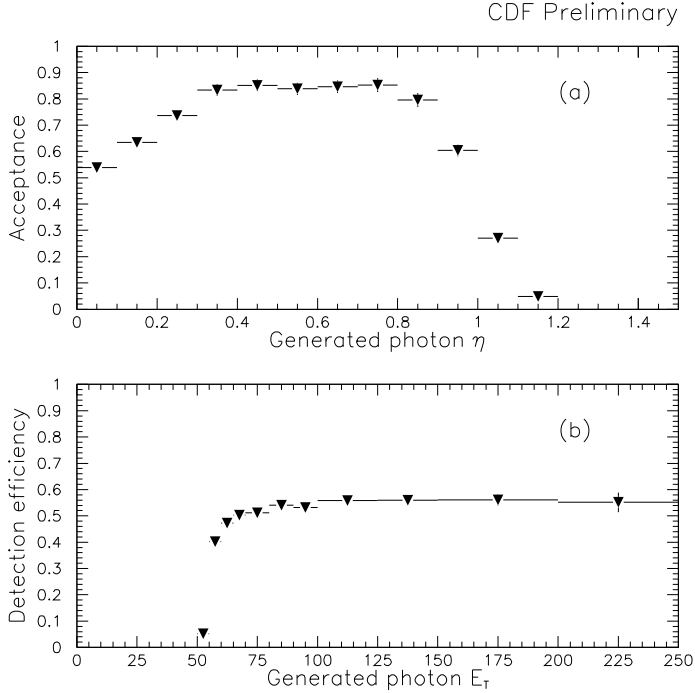,height=10cm}
	\caption{\it Detector acceptance and detection efficiency of
	photons at CDF.}
	\label{fig:gamma_met_acc_eff}
\end{figure}

Since no signal is seen in our data, these 11 events can be used to
exclude models or to set limits on models. In order to do that, a
generic detector acceptance and efficiency for photons has to be
measured. They are shown in Figure~\ref{fig:gamma_met_acc_eff}. We can
then convolute photon + \MET\ events from any model with the acceptance,
efficiency, and detector resolution, to calculate the total event
acceptance for the model.

We look into two models. The first one is a superlight
gravitino\cite{gravitino} model in which the only supersymmetric
particle light enough to be produced at Tevatron is the gravitino
$\widetilde{G}$. The process is $q\overline{q} \rightarrow
\widetilde{G}\widetilde{G}\gamma$, in which the photon comes from
initial state radiation and serves as a tagger for the process. The {\sc
SUSY}-breaking scale $F$ can be related to the mass of the gravitino
$m_{3/2}$ by $F=\sqrt{3}m_{3/2}M_{pl}$, where $M_{pl}
\approx 2.4 \times 10^{18}$ is the Planck scale. By convoluting the model
through our Monte Carlo and acceptance, efficiency, and resolution
functions, we obtain a new\footnote{The past best 95\% C.L. limit has
been $\sqrt{F}>217 \; {\rm GeV}$ from CDF
jet+\MET\cite{cdf_jet_met_gravitino}.} 95\% C.L. limit of $\sqrt{F} >
221 \; {\rm GeV}$, or $m_{3/2} > 1.17 \times 10^{-5} \; {\rm eV}$.

The second model we look into is large extra dimension (LED) in the
framework of Arkani-Hamed, Dimopoulos, and Dvali\cite{ADD}. The relevant
process is $q\overline{q} \rightarrow G_{kk}\gamma$. We list the
calculated 95\% C.L. limit on the effective Planck scale $M_{D}$ in the
GRW\cite{GRW} convention in Table~\ref{tbl:gamma_met_led_limit}. Also
listed in the table is the best results obtained by the LEP
experiments. As in the monojet case, the CDF limit extends that from the
LEP at higher extra dimension due to the higher center-of-mass energy of
the Tevatron.

\begin{table} [h!tb]
\centering
\caption{\it Lower limits in the effective Planck scale $M_{D}$ as a
function of number of extra dimension $n$ obtained by CDF search for LED in
photon + \MET\ channel.}
\vskip 0.1 in
\begin{tabular}{|c|c|c|c|} \hline
$n$ & 4 & 6 & 8 \\ \hline \hline
$M_{D}$ (GeV) & 549 & 581 & 602 \\
$M_{D}$ (GeV) (LEP) & 680\cite{delphi_led} & 510\cite{delphi_led} &
411\cite{l3_led} \\
\hline
\end{tabular}
\label{tbl:gamma_met_led_limit}
\end{table}

\subsection{Quaero -- automatic data analysis machine at D\O\cite{quaero}}
Quaero, which means ``to seek'' in Latin, uses D\O\ data which are
publically available on the internet and automatically optimizes
searches for any signal provided by the user. The data sets are
categorized according to their final states. The backgrounds and their
respective fractions have been calculated and are available to the
user. It is understood that the data are well explained by the expected
background and that the goal of the analysis is to set $\sigma^{95\%}$,
the 95\% C.L. upper limit, on the cross section of the model. At the
Quaero web page, http://quaero.fnal.gov, users can use {\sc
PYTHIA}\cite{pythia} to generate their model events which results in one
of the available final states. They can then define a variable set
$\vec{x}$ to be used to optimize the search. The optimization algorithm
has the following steps:

\begin{itemize}
\item Kernel density estimation\cite{kernel} is used to obtain a signal
and background probability distribution $p(\vec{x}|s)$ and $p(\vec{x}|b)$,
respectively;
\item A discriminant function is defined as\cite{kernel}:
\begin{equation}
D(\vec{x}) = \frac{p(\vec{x}|s)}{p(\vec{x}|s)+p(\vec{x}|b)} \nonumber;
\end{equation}
\item The sensitivity $S$ is defined as the reciprocal of
$\overline{\sigma^{95\%}}$, which is the 95\% C.L. limit on model cross
section as a function of $D_{cut}$ on $D(\vec{x})$. An optimal $D_{cut}$
on $D(\vec{x})$ is chosen to minimize $\overline{\sigma^{95\%}}$, thus
maximize $S$;
\item The region of variable space having $D(\vec{x}) > D_{cut}$ is used
to determine the actual 95\% C.L. cross section upper limit
$\sigma^{95\%}$.
\end{itemize}

Based on Run 1 data, Quaero has set $\sigma^{95\%}$ for various models,
including SM higgs production: $h \rightarrow WW \rightarrow e\mbox{\MET}2j$, $h
\rightarrow ZZ \rightarrow ee2j$, $Wh \rightarrow e\mbox{\MET}2j$, and
$Zh \rightarrow ee2j$; $W'$ and $Z'$ production: $W' \rightarrow WZ
\rightarrow e\mbox{\MET}2j$, $Z' \rightarrow t\overline{t} \rightarrow
e\mbox{\MET}4j$; and leptoquark production: $LQ\overline{LQ} \rightarrow
ee2j$.

\section{Conclusion}
We have presented in this paper nine analyses which were finalized in
year 2001. Even though we did not observe any signature of new physics,
we are able to set stronger limits on them. New tools and techniques
have been developed to equip us for more challenging searches. With more
powerful detectors for both CDF and D\O, and an order of magnitude of
increase in luminosity for expected for Run 2, we are looking forward to
more exciting searches and possibly discoveries.

\section{Acknowledgment}
The author is thankful to both CDF and D\O\ Collaborations for the
opportunity to present these new results at La~Thuile 2002'. In
particular the author would like to thank the following persons who
actually carried out the analyses and provided the detail information:
Abdelouahab~Abdesselam, Sudeshna~Banerjee, and Hai~Zheng of the D\O\
Collaboration, and David~Gerdes, Chris~Hays, Teruki~Kamon, Minjeong~Kim,
Bruce~Knuteson, Yoshiyuki~Miyazaki, Simona~Murgia, Peter~Onyisi, and
Masashi~Tanaka of the CDF Collaboration.


\begin{thebibliography}{99}

\bibitem{cdfnim} CDF Collaboration, F.~Abe \etal,
Nucl.~Instr.~Methods~Phys.~Res.~A~{\bf 271}, 387 (1988).

\bibitem{d0nim} D\O\ Collaboration, S. Abachi \etal,
Nucl.~Instr.~Methods~Phys.~Res.~A~{\bf 338}, 185 (1994).

\bibitem{msugra} For reviews see P.~Nath, R.~Arnowitt, and
A.~H.~Chamseddine, ``Applied $N=1$ Supergravity'' (World Scientific,
Singapore, 1984); H.~P.~Nilles, Phys.~Rept.~{\bf 110},~1~(1984).

\bibitem{susy} D.~V.~Volkov and V.~P.~Akulov, Phys.~Rev.~Lett.~{\bf
46B},~109~(1973); J.~Wess and B.~Zumino, Nucl.~Phys.~{\bf
B70},~39~(1974).

\bibitem{r-parity} G.~R.~Farrar and P.~Fayet, Phys. Lett. B {\bf 76},
575 (1978).

\bibitem{mssm} For a review, see e.g., H.~Haber and G.~Kane,
Phys.~Rept.~{\bf 117},~75~(1985).

\bibitem{rpv_stop_aleph} ALEPH Collaboration, hep-ex/0011008.

\bibitem{rpv_2mu4jets} D\O\ Collaboration, V.~M.~Abazov \etal, hep-ex/0111053.

\bibitem{se_msugra} D\O\ Collaboration, V.~M.~Abazov \etal, hep-ex/0205002.

\bibitem{spythia} S.~Mrenna, hep-ph/9609360.

\bibitem{lepsusy} LEP2 {\sc SUSY} Working Group, http://www.cern.ch/LEPSUSY.

\bibitem{cdf_jets_met_msugra} CDF Collaboration, T.~Affolder \etal,
Phys. Rev. Lett.~{\bf 88}, 041801 (2002).

\bibitem{ADD} N.~Arkani-Hamed, S.~Dimopoulos, and G.~Dvali,
Phys.~Lett.~B~{\bf 429}, 263 (1998).

\bibitem{GRW} G.~F.~Giudice, R.~Rattazzi, and J.~D.~Wells,
Nucl.~Phys.~B~{\bf 544}, 3 (1999).

\bibitem{d0_led_2e_2g} D\O\ Collaboration, B.~Abbott \etal,
Phys.~Rev.~Lett.~{\bf 86}, 1156 (2001).

\bibitem{lykken_matchev} J.~Lykken and C.~Matchev, private code.

\bibitem{pythia} M.~Bengtsson and T.~Sj\"{o}strand,
Comp.~Phys.~Comm.~{\bf 43}, 367 (1987).

\bibitem{lq_jets_met} D\O\ Collaboration, V.~M.~Abazov \etal,
Phys.~Rev.~Lett.~{\bf 88}, 191801 (2002).

\bibitem{lq_scalar} P.~H.~Frampton, Mod.~Phys.~Lett.~A~{\bf 7}, 559 (1992).

\bibitem{lq_vector} H.~Georgi and S.~Glashow, Phys.~Rev.~Lett.~{\bf 32},
438 (1974); J~.C~.Pati and A.~Salam, Phys.~Rev.~D~{\bf 10}, 275 (1974).

\bibitem{lq_fcnc_constraint} H.~U.~Bengtsson, W.~S.~Hou, A.~Soni, and
D.~H.~Stork, Phys.~Rev.~Lett.~{\bf 55}, 2762 (1985).

\bibitem{lq_vector_couplings} J.~Bl\"{u}mlein, E.~Boos, and A.~Kryukov,
Z.~Phys.~C~{\bf 76}, 137 (1997).

\bibitem{d0_lq_run1a} D\O\ Collaboration, B.~Abbott \etal,
Phys.~Rev.~Lett.~{\bf 80}, 2051 (1998); D\O\ Collaboration, V.~M.~Abazov
\etal, hex-ex/0105072.

\bibitem{d0_lq_1gen} D\O\ Collaboration, V.~M.~Abazov \etal,
Phys.~Rev.~D~{\bf 64}, 092004 (2001).

\bibitem{d0_lq_2gen} D\O\ Collaboration, B.~Abbott \etal,
Phys.~Rev.~Lett.~{\bf 83}, 2896 (1999); D\O\ Collaboration, B.~Abbott
\etal, Phys.~Rev.~Lett.~{\bf 84}, 2088 (2000).

\bibitem{gamma_met} CDF Collaboration, D.~Acosta \etal, hep-ex/0205057.

\bibitem{gravitino} A.~Brignole, F.~Feruglio, M.~L.~Mangano, and
F.~Zwirner, Nucl.~Phys.~B {\bf 526}, 136 (1998); erratum-ibid.~B~{\bf
582}, 759 (2000).

\bibitem{cdf_jet_met_gravitino} CDF Collaboration, T.~Affolder \etal,
Phys.~Rev.~Lett.~{\bf 85}, 1378 (2000).

\bibitem{delphi_led} P.~Abreu \etal, Eur.~Phys.~J.~C~{\bf 17}, 53 (2000).

\bibitem{l3_led} M.~Acciarri \etal, Phys.~Lett.~B~{\bf 470}, 268 (1999).

\bibitem{quaero} D\O\ Collaboration, B.~Abbott \etal,
Phys.~Rev.~Lett.~{\bf 87}, 231801 (2001); http://quaero.fnal.gov/quaero/~.

\bibitem{kernel} L.~Holmstrom, S.~Sain, and H.~Miettinen,
Comp.~Phys.~Commun.~{\bf 88}, 195 (1995).

\end{thebibliography}
\end{document}